%% file: main.tex
\documentclass[sigconf,nonacm]{acmart}

\usepackage{graphicx}
\usepackage{amsmath}
\usepackage{booktabs}
\usepackage{multirow}
\usepackage{hyperref}
\usepackage{subcaption}
\usepackage{tabularx}
\usepackage{pifont}
\usepackage{makecell}
\usepackage{enumitem}
\usepackage{footnote}
\usepackage{breakcites}

\usepackage{anyfontsize}
\makesavenoteenv{tabular}
\makesavenoteenv{table}

\newcommand{\reproducible}{{$\CIRCLE$}}%
\newcommand{\unreproducible}{{$\Circle$}}%
\newcommand{\partlyreproducible}{$\LEFTcircle$}
\newcommand{\notapplic}{---}

\newcommand{\cmark}{\ding{51}}%
\newcommand{\xmark}{\ding{55}}%
\usepackage{wasysym}
\usepackage{tikz}
\usepackage{collcell}
\usepackage[table]{colortbl}

\newcommand*{\MinNumber}{11.5}%
\newcommand*{\MidNumber}{13} %
\newcommand*{\MaxNumber}{33.5}%

\newcommand{\ApplyGradient}[1]{%
        \ifdim #1 pt > \MidNumber pt
            \pgfmathsetmacro{\PercentColor}{max(min(100.0*(#1 - \MidNumber)/(\MaxNumber-\MidNumber),100.0),0.00)} %
            \hspace{-0.66em}\edef\x{\noexpand\cellcolor{red!\PercentColor!yellow} }
            \x #1
        \else
            \pgfmathsetmacro{\PercentColor}{max(min(100.0*(\MidNumber - #1)/(\MidNumber-\MinNumber),100.0),0.00)} \hspace{-0.66em}\edef\x{\noexpand\cellcolor{green!\PercentColor!yellow} }
            \x #1
        \fi
}

\newcolumntype{R}{>{\collectcell\ApplyGradient}c<{\endcollectcell}}

\newcommand*{\MinNumberTwo}{6.28}%
\newcommand*{\MidNumberTwo}{6.5} %
\newcommand*{\MaxNumberTwo}{12.5}%

\newcommand{\ApplyGradientTwo}[1]{%
        \ifdim #1 pt > \MidNumberTwo pt
            \pgfmathsetmacro{\PercentColor}{max(min(100.0*(#1 - \MidNumberTwo)/(\MaxNumberTwo-\MidNumberTwo),100.0),0.00)} %
            \hspace{-0.66em}\edef\x{\noexpand\cellcolor{red!\PercentColor!yellow} }
            \x #1
        \else
            \pgfmathsetmacro{\PercentColor}{max(min(100.0*(\MidNumberTwo - #1)/(\MidNumberTwo-\MinNumberTwo),100.0),0.00)} \hspace{-0.66em}\edef\x{\noexpand\cellcolor{green!\PercentColor!yellow} }
            \x #1
        \fi
}

\newcolumntype{G}{>{\collectcell\ApplyGradientTwo}c<{\endcollectcell}}

\AtBeginDocument{%
  \providecommand\BibTeX{{%
    \normalfont B\kern-0.5em{\scshape i\kern-0.25em b}\kern-0.8em\TeX}}}

\acmDOI{}

\acmConference[ACSAC '22]{Annual Computer Security Applications Conference}{December 05--09,2022}{Austin, Texas, USA}
\acmBooktitle{Proceedings of the 2022 Annual Computer Security Applications Conference (ACSAC '22), December 05--09, Austin, Texas, USA}
\acmISBN{}

\begin{document}

\title{Techniques for Continuous Touch-Based Authentication Modeling}


\author{Martin Georgiev}
\affiliation{%
  \institution{University of Oxford} \city{Oxford} \country{UK}
  }
\email{martin.georgiev@cs.ox.ac.uk}
  
\author{Simon Eberz}
\affiliation{%
  \institution{University of Oxford} \city{Oxford} \country{UK}
  }
\email{simon.eberz@cs.ox.ac.uk}

\author{Ivan Martinovic}
\affiliation{%
    \institution{University of Oxford}
    \city{Oxford}
    \country{UK}
  }
\email{ivan.martinovic@cs.ox.ac.uk}


\begin{abstract}
The field of touch-based authentication has been rapidly developing over the last decade, creating a fragmented and difficult-to-navigate area for researchers and application developers alike due to the variety of methods investigated.
In this study, we perform a systematic literature analysis of 30 studies on the techniques used for feature extraction, classification, and aggregation in touch-based authentication systems as well as the performance metrics reported by each study.
Based on our findings, we design a set of experiments to compare the performance of the most frequently used techniques in the field under clearly defined conditions.
In addition, we introduce three new techniques for touch-based authentication: an expanded feature set (consisting of 149 unique features), a multi-algorithm ensemble-based classifier, and a Recurrent Neural Network based stacking aggregation method.
The comparison includes 14 feature sets, 11 classifiers, and 5 aggregation methods.
In total, 219 model configurations are examined and we show that our novel techniques outperform the current state-of-the-art in each category.
The results are also validated across three different publicly available datasets.
Finally, we discuss the findings of our investigation with the aim of making the field more understandable and accessible for researchers and practitioners.
\end{abstract}







\maketitle

\section{Introduction}
Over the past two decades, the world has become increasingly reliant on mobile devices, such as smartphones and tablets, for professional, social, and leisure activities. 
It is estimated that 81\% of the world population is in possession of a smartphone today, including children and the elderly \cite{ericsson}. 
Mobile devices have grown to deliver services far beyond that which can be considered trivial and mundane and now are used for activities that involve the processing and storage of sensitive and private information. 

Almost all mobile devices offer basic mechanisms for authentication - proving ownership and identity, to prevent unauthorized people from accessing private data. 
Thus far, the prevalent methods for authentication used by mainstream smartphones include pins, pattern unlocking, face recognition, and fingerprint matching. 
However, such methods have been shown to not always be reliant as they can be inconvenient to use \cite{ready-to-lock} or prone to various attacks \cite{smudge-attack,fingerprint-attack-1, face-recognition-attack}.

More recently, the scientific community has introduced continuous mobile authentication systems which ensure user identity throughout the whole session.
Such systems can make use of the unique way people type on their phone \cite{cont-typing}, the sensors of their smartphones \cite{hmog-sensors}, by continuously recording video of their faces \cite{cont-face} and other behavioral characteristics.
Touch-based authentication methods, specifically, make use of the way people interact with the screen of their smartphones. 
These methods rely on assessing whether the interaction patterns of the person authenticating match the general behavior of the original owner.
Touch-based authentication offers significant advantages over traditional methods, as it is relying on dynamic and inherent properties of users rather than static attributes and memorized sequences.
Touch-based authentication systems can be used as a second-factor authentication in sensitive applications such as banking and finance, where suspicious behavior can be flagged and additional checks requested for certain transactions.
The technology can also be used to detect whenever a malicious user gets a hold of an unlocked phone. 
The system, then, would detect the difference in behavior and request an additional authentication method to verify identity. 
However, despite almost a decade of research into this field and positive sentiment from the community for such technology \cite{continuous-support}, touch-based authentication methods still lack widespread adoption and integration into our daily devices. 
This issue can be attributed to the way studies are evaluated and the resulting overestimation of performance \cite{common-evaluation}. 
However, contributions in this area are also severely fragmented, primarily due to a lack of a common framework to compare and evaluate models against a well-defined benchmark to accurately assess what can be considered state-of-the-art. 
In order to make impactful contributions and improve techniques for touch-based authentication, it is imperative to clarify the landscape and provide methods to reason about model performance. 
\input{figures/lifecycle} 

To this end, in this paper, we aim to answer the following research questions:

\begin{enumerate}
    \item RQ1: What are the current techniques for performing touch-based authentication? Which features, classifiers, aggregation methods, and metrics are used and how can we derive a theory for grouping them into common categories?
    \item RQ2: How can we establish the best-performing techniques despite the variety in models and evaluation datasets?
    \item RQ3: Which techniques are most important in building robust and well-performing touch-based authentication methods and how can we use this to improve upon the state-of-the-art in the field?
\end{enumerate}

To answer our research questions, we make use of a two-fold approach. 
First, we conduct a systematic literature review to establish a broad understanding of existing methods for touch-based authentication. 
We then extract features, classifiers, aggregations, and metrics used for each study and categorize feature extraction and aggregation methods. 
Second, to determine and improve on state-of-the-art models, we evaluate a carefully selected range of models along common parameters using three open-source datasets, allowing us to understand which are the best-performing methods. 
We use this to propose a set of techniques that outperform the current state-of-the-art and evaluate them against the current best-performing methods.

In this paper we make the following contributions:
\begin{itemize}
    \item Through a systematic literature review we analyzed 30 papers, extracted 149 features, and categorized touch-based authentication methods.
    \item We performed a comparison of 219 model configurations across 3 datasets, allowing us to determine the best-performing features, classifiers, and aggregation methods in the field.
    \item We proposed a novel set of accumulated features, an ensemble-based classification model, and a Recurrent Neural Network (RNN) stacking-based aggregation method, all of which outperform the current state-of-the-art.
\end{itemize}

\section{Background}

Continuous mobile authentication systems passively verify that a user enrolled in a device is the one persisting on it.
This is done by comparing new patterns of interaction with the legitimate ones of the enrolled user.
When a significant mismatch is detected, the system can block the malicious user and notify the owner of the device.
For instance, this can be useful when a pin has been stolen or an unauthorized user gets access to an unlocked phone.
Then, when the unauthorized user starts using the phone, they will be stopped by the continuous authentication system as the pattern of usage deviates from the owner of the smartphone.

There are many ways in which continuous authentication on mobile devices can be performed, including keystrokes \cite{keystroke-mobile}, taps \cite{tapping}, multi-touch gestures \cite{multitouch-only}, sensors \cite{only-sensors,hmog-sensors}, freeform gestures \cite{free-form-gestures}, active vibration signal \cite{vibration} and active gesture methods \cite{gestures-mturk}.
However, the focus of this paper is on touch-dynamics - horizontal and vertical displacements on touch-capacitive displays done using a single finger which are called strokes. 
These are derived from the coordinates and pressure points at contact while interacting with the screen.
Some touch-based authentication systems augment these strokes with additional data such as sensor information from the accelerometer and gyroscope.
However, the focus of our study is on stroke-based systems.

The lifecycle of a continuous touch-based authentication system is illustrated in Figure~\ref{fig:lifecycle}.
The data collection step could be the experimental setup for a study or in the case of a deployed system it could be the enrollment phase where individual templates of behavior are created. 
The feature extraction step in touch-based authentication aims at obtaining unique information from touchscreen interactive sessions with the smartphone which can be used to differentiate between users of the system.
The classifier step relies on models to make a decision about the legitimacy of a particular swipe based on enrollment patterns.
These are typically machine learning algorithms that are trained on the features extracted in the previous steps.
Furthermore, a single swipe may not provide enough distinguishing information for an acceptable authentication performance.
For this purpose, some systems perform aggregation of successive swipes to improve system performance.
In the final step, a variety of metrics could be used to capture and report the success of the biometric system.

\subsection{Related Work}

The first touch-based mobile authentication systems were proposed in the early 2010s inspired by previous work on the use of mouse movement patterns for authentication on desktop computers \cite{mouse-movement} and other continuous authentication developments.
Several studies \cite{overview-1,overview-2,overview-3,overview-4} survey the historical development of touch-based authentication, reporting on the progress, remaining challenges, and performance of models in the field.
However, such surveys do not consider the quantitative performance differences between feature extraction, classification and aggregation methods under fair conditions, thus lacking best-practice recommendations for researchers and practitioners.
Other works \cite{benchmark-touch, which-verifiers-work} do perform a comparison between a limited number of classification methods.
Nevertheless, our study differentiates itself by its magnitude in terms of classifiers examined and investigation of numerous feature extraction and aggregation methods.
Furthermore, we use the accumulated knowledge to propose methods that perform better in each step of the system's lifecycle.

\subsection{Datasets}

There are dozens of studies that design and implement their own experiments for data collection as shown in \cite{common-evaluation}.
However, the vast majority do not share the resulting dataset upon publication.
We present 9 publicly available touch-based authentication datasets in Table~\ref{tab:datasets}.
We use the following criteria derived from \cite{common-evaluation} to select the datasets applicable to our investigation.
Naturally, the data in question needs to be accessible at the time of requesting it.
That is not always the case as some of the servers hosting the data have gone down since paper publication.
The dataset itself should contain a group of users using the same smartphone model.
The users should have performed at least 2 separate sessions of the experimental tasks.
Furthermore, each swipe needs to contain information about its (X, Y) coordinates as well as touch area and pressure values.
These are required for the majority of feature extraction approaches.
The only publicly available datasets which we consider usable under these conditions are Touchalytics \cite{touchalytics}, Bioident \cite{information-revealed} and CEP \cite{common-evaluation}. 
We focus on these three datasets in the rest of this study and describe the reason for not using the other datasets in the ``Notes" column in Table~\ref{tab:datasets}. 
For each of the three datasets, we select the largest subset of users who use the same phone model and perform 2 or more sessions.

\input{tables/datasets}



\section{Techniques for touch-based authentication }
In this section, we perform a systematic literature review of papers proposing systems for touch-based authentication. 
We quantify the prevalence of techniques for feature extraction, classification and aggregation in touch-based authentication systems as well as the methods for measuring model performance. 
Furthermore, we group the approaches into semantically similar categories in order to consolidate the understanding of the field.


\subsection{Methods}

The objective of the literature review is to understand the methods used in each of the core components of the continuous touch-based authentication lifecycle (Figure~\ref{fig:lifecycle}) so that we can next re-evaluate them in a common benchmark. 
For our systematic literature review we relied on PRISMA \cite{moher2009preferred} to guide the search strategy to identify articles that proposed and evaluated touch-based authentication models. The search was limited to English language and peer-reviewed published articles. We exclusively made use of the Google Scholar database and used the following search terms: ((\textit{touch-based} OR \textit{touchscreen}) AND (\textit{authentication} OR \textit{biometric*})) OR \textit{touch dynamics} OR \textit{touch biometrics} OR \textit{touch authentication} OR \textit{continuous touch}.

Articles were included if the methods focus on touch-based authentication and made use of machine-learning-based models. 
We did include articles that, in addition to touch-dynamics-based features, also make use of other features, such as ones based on accelerometer and gyroscope data.
However, in our performance evaluation, we do not make use of the additional features as such data is not available in all the publicly available datasets we consider in our study, making comparison difficult.
We then implemented an ancestry approach with the articles meeting the inclusion and exclusion criteria.
Our keyword-based search identified a total of 685 articles. Following the screening step, we were left with 103 articles.
Screening involved the inspection of the title and description of papers for eligibility, as well as removing any duplicates.
After our complete eligibility criteria were applied, we included 30 articles in the final review. 
For each article, we manually tabulated the features, classification methods, and aggregation methods, as well as the metrics used to evaluate performance.



\input{tables/studies}

\subsection{Findings}

The results of our survey are summarized in Table~\ref{tab:papers}.
We organize our findings into four sections, each encapsulating the results for a step of the touch-based authentication lifecycle - features, classifiers, aggregations, and metrics (reporting of results).
We deliberately do not include the performance reported by each study due to the variety of metrics and datasets used, making the exact data meaningless to compare across the studies.


\subsubsection{Features}

We found that we can broadly categorize features according to three classes:

\begin{itemize}
    \item \textit{Swipe-based}: These features are based on data derived from individual swipes. Typically, the features are generated by examining a list of (X, Y, pressure, area) points that form a complete swipe.
    Examples of such features include the starting X or Y position, the length of the swipe trajectory, the average pressure of the swipe, etc.
    \item \textit{Image-based}: These methods are based on generating an image that represents the swipe on a 2D plane.
    The images are then fed into image processing pipelines for texture and shape extraction \cite{trace-maps} or to compute a difference score between images \cite{statistical-images}.
    \item \textit{Session-based}: These methods are based on the properties of whole sessions, rather than a single or small group of swipe-based features.
    Examples of such features include the number of swipes per session, the average time duration of swipes per session, the average time duration between swipes per session, etc.
\end{itemize}

Most studies make use of \textit{swipe-based} features (80\%).
13\% make use of \textit{session-based} features and 7\% of \textit{image-based} ones. 
The prevalence of \textit{swipe-based} features can be explained by the high computational cost associated with image processing and the long feature accumulation period of \textit{session-based} methods, during which the device is left unprotected.

In total, 16 (67\%) of the 24 \textit{swipe-based} based studies, defined their features and extraction methods sufficiently to be reproducible.
Another 4 (17\%) of the \textit{swipe-based} studies have feature sets that can be only partially reproduced as a number of features do not have clear and non-ambiguous descriptions.
For instance, one of the articles has a feature described as ``the angle of moving during swiping" \cite{towards-cont-passive}, without detailing how and which angle is calculated.  
For further 4 (17\%) of the \textit{swipe-based} studies, we could not infer the individual features used for authentication due to broad category definitions rather than specific feature descriptions or the information not being provided by the authors at all. 

In total we identified 149 \textit{swipe-based} features from all papers. 
The list of features can be seen in Appendix~\ref{sec:all_features}. 
The average number of features per paper is 24 where the largest number of \textit{swipe-based} features identified in a single paper is 59 \cite{towards-cont-passive}, and the smallest is 5 \cite{fusing-swiping}. 

In this study, we focus on the geometric feature extraction approach as it is implemented by 83\% of studies examined in Table~\ref{tab:papers}.
Furthermore, we argue that it is the most realistic approach given the computational, time, and security constraints of a continuous mobile authentication system.

\footnotetext[1]{\href{http://www.mariofrank.net/touchalytics/}{http://www.mariofrank.net/touchalytics/}} 
\footnotetext[2]{\href{http://www2.latech.edu/ phoha/BTAS-2013.htm}{http://www2.latech.edu/ phoha/BTAS-2013.htm}} 
\footnotetext[3]{\href{http://www.cudroid.com/urmajesty}{http://www.cudroid.com/urmajesty}}
\footnotetext[4]{\href{https://umdaa02.github.io/}{https://umdaa02.github.io/}}
\footnotetext[5]{\href{https://ms.sapientia.ro/~manyi/bioident.html}{https://ms.sapientia.ro/~manyi/bioident.html}}
\footnotetext[6]{\href{http://zasyed.com/jss18dataset.html}{http://zasyed.com/jss18dataset.html}}
\footnotetext[7]{\href{https://zenodo.org/record/2598135}{https://zenodo.org/record/2598135}}
\footnotetext[8]{\href{https://github.com/BiDAlab/HuMIdb}{https://github.com/BiDAlab/HuMIdb}}
\footnotetext[9]{\href{https://ora.ox.ac.uk/objects/uuid:5f1abaa7-52a4-430b-9208-128d9f1832fd}{https://ora.ox.ac.uk/objects/uuid:5f1abaa7-52a4-430b-9208-128d9f1832fd}}

\subsubsection{Classification}
\label{sec:prev_class}

\begin{figure*}[!t]
\includegraphics[width=1.00\linewidth]{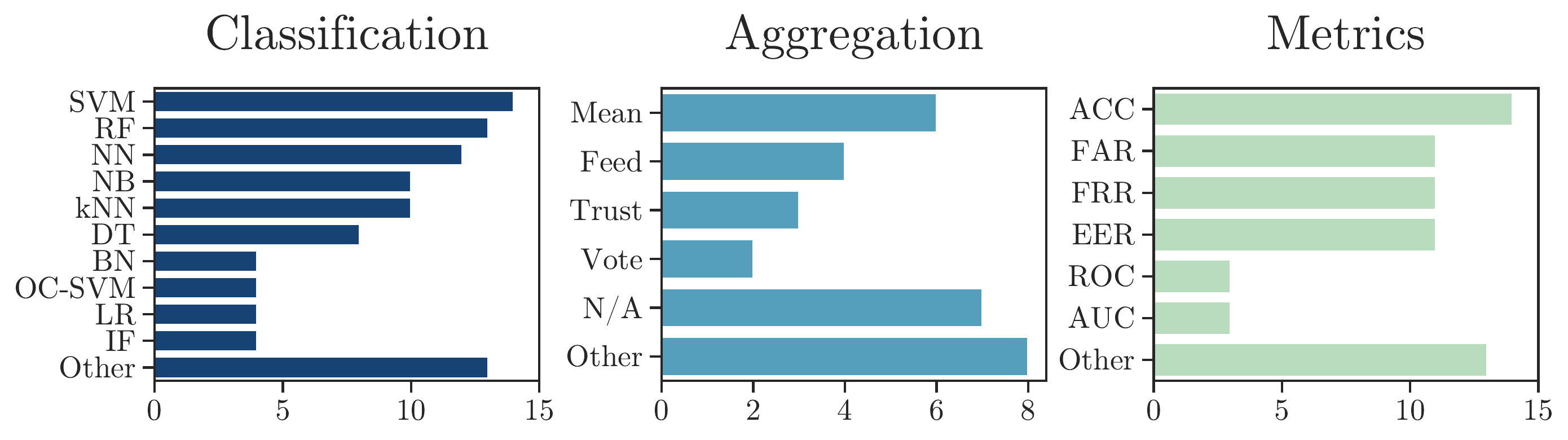}
\caption{The prevalence of classifiers, aggregation methods, and performance metrics in touch-based authentication studies. The ``Other" category means the particular methods have been used less than 3 times in the case of Classification and Metrics and less than 2 times in the case of Aggregation.}
\vspace{6px}
\label{fig:statistics}
\end{figure*}

The studies examined in Table~\ref{tab:papers} include a total of 27 unique classification approaches, many of which are available as out-of-the-box implementations in standard machine learning libraries.
We present the prevalence of the most frequently used classification models in Figure~\ref{fig:statistics}.
The most common are Support Vector Machine, Random Forest, and Neural Networks with 47\%, 43\% and 40\% of studies including them as part of their analysis respectively.
The maximum number of classifiers included in a single study is 12 \cite{one-class}.
In total, 12 (44\%) of the classifiers appear only once across all studies.

\subsubsection{Aggregation}
\label{sec:prev_agg}

In total, we found that a large proportion of the studies (77\%) perform the optional aggregation step in their touch-based authentication systems.
There are multiple ways of approaching the processing of a group of swipes to extract optimal performance. 
We found that we can broadly categorize the aggregation methods into the following four classes:

\begin{itemize}
    \item \textit{Mean}: The average or median value of the scores of each swipe returned by the classifier.
    \item \textit{Vote}: The most common binary prediction (legitimate user or attacker) for each swipe decided by the classifier.
    \item \textit{Feed}: In this approach, all of the stroke features are fed into the model at once and a single prediction is obtained.
    For instance, if there are 10 features with a window size of 5 (i.e. group of five consecutive swipes), we input all 50 features into the model at once.
    \item \textit{Trust}: There is a large variation in this category but the general idea is to make use of a statistical formula that outputs a score as new swipes are considered.
    The score is updated by rewarding positive predictions and penalizing negative predictions proportionally to the individual classifier predictions. 
    An instance of such aggregation methods is the dynamic trust model \cite{trust-model}, which is tailored to continuous authentication biometric systems. 
    This specific implementation has been used in \cite{swipe-gesture} and \cite{behave-sense}.
\end{itemize}


We present the prevalence of each of these methods in Figure~\ref{fig:statistics}.
The \textit{Mean} aggregation approach is the most frequently used one (20\%).
The \textit{Vote}, \textit{Feed}, \textit{Trust} methods are used in 6\%, 13\% and 10\% of the studies respectively.
As mentioned, 23\% of the studies do not use aggregation at all, and a further 27\% use solutions that do not fall into the categories described above.
For instance, the systems using session-based features are making decisions based on a large aggregation of swipes but cannot be included in any of the other categories we describe.

\subsubsection{Metrics}

Depending on the needs of a particular system, there are a variety of metrics that can be used to measure the performance of a model for touch-based authentication. 
These include FAR (False Acceptance Rate), FRR (False Rejection Rate), EER (Equal Error Rate), Accuracy, ROC curve (Receiver Operating Characteristic), and others.
Statistics for the prevalence of these metrics in the field can be found in Figure~\ref{fig:statistics}.
The variety of metrics shown illustrates the difficulty in comparing results reported in touch-based authentication studies.
In this study, we aim to ease this comparison by reporting on differences in approaches when they are examined under the same conditions and by reporting the results using the same metrics.
The reporting of results and impact of metric choices for authentication systems has been discussed thoroughly by Sugrim et al. \cite{robust-performance}.

In this paper, we report our results using the EER metric.
The EER is the point at which the FAR and FRR are equal on the ROC curve.
The ROC curve is obtained by varying the threshold for acceptance into the biometric system.
Therefore, there is a value of the threshold which corresponds to the EER.
While some systems might benefit from choosing thresholds for optimizing better FAR or FRR we believe EER is the most representative of the general performance of a system.
This is also supported by the related work \cite{robust-performance,eberz,common-evaluation}.
In particular, \cite{common-evaluation} show that when comparing two touch-based authentication models, the performance differences between them on the ROC curve are largely consistent with the difference at the EER point.


%


\section{Performance evaluation}
\label{sec:experiments}

In this section, we evaluate touch-based authentication techniques, determine which are the best-performing ones, and introduce a new state-of-the-art feature set, classifier, and aggregation method.

\subsection{Methods}
\label{sec:eval_methods}

The objective of this performance evaluation is to determine the best-performing existing feature sets, classifiers, and aggregation methods.
Furthermore, we aim to identify a set of novel techniques and compare them to the current state-of-the-art. 
Finally, the study aims to understand whether the results obtained are valid across multiple publicly available datasets.
To this end, we examine how each classifier performs on different feature sets and then compare aggregations methods independently.

Throughout our study, we follow the recommendations from \cite{common-evaluation} for fair evaluation of touch-based authentication systems.
We create the following model for each user and record the mean EER of all users at the end.
We select users which have performed at least two sessions and use the same phone model.
At first, we split the data of a target user, selecting their first 80\% of sessions for positive training data and the remaining 20\% for positive testing data.
We split the rest of the users into independent training and testing groups at random.
The users in each group never overlap.
The negative data for training or testing is then obtained by selecting a swipe at random while cycling through the respective group of users until the number of negative training or testing swipes is equal to the positive one.
The combined training set is then used to train a binary model and the testing set for evaluating the performance of the model.
This whole process is repeated 10 times for each experiment and we report the mean of the results from each repetition. 
At each of these iterations, we randomly select the training and testing user groups.
The one-class classifiers employ the same process, however, the negative training data is not used.

The SVM, RF, NB, kNN, DT, OC-SVM, LR, and IF classifiers we investigated were implemented using the popular \texttt{scikit-learn}~\cite{sklearn_api} machine learning library.
The Neural Networks were implemented using \texttt{Tensorflow} \cite{tensorflow} and the \texttt{Keras} \cite{keras} API.
The Bayesian Network implementation was done on the WEKA \cite{weka} machine learning library using a Python wrapper.

The implementation details of each classification algorithm were left as close to the default as possible.
Where we had to make decisions (e.g., in the case of kNN and Neural Networks), we looked at the related work and performed preliminary experiments to decide on the hyperparameters.
The final parameters for each classifier are given below:

\begin{itemize}
    \item Support Vector Machine (SVM) - RBF kernel with a 'scale' coefficient and probability estimations enabled.
    \item Random Forests (RF) - 100 estimators, max depth of 20, and probability estimations enabled.
    \item Neural Network (NN) - feed-forward with three hidden layers of 150, 150, and 75 with a 'ReLU' activation function.
    The output layer has a 'Sigmoid' activation function which outputs a probability of a match between 0 and 1.
    Batch-normalization is applied at each layer and a 0.3 dropout between the hidden layers. 
    The optimizer is 'Adam' with a 'binary cross-entropy' loss function. The network is trained with a batch size of 20 over 50 epochs.
    \item Naive Bayes (NB) - gaussian naive bayes implementation.
    \item k Nearest Neighbors (kNN) - number of neighbors - 18.
    \item Decision Trees (DT) - gini criterion and no maximum depth.
    \item Bayesian Network (BN) - K2 algorithm for structure learning and Simple Estimator for predictions.
    \item One-Class - Support Vector Machine (OC-SVM) - RBF kernel with a 'scale' coefficient.
    \item Logistic Regression (LR) - LBFGS solver with L2 penalty and maximum iterations of 1000.
    \item Isolation Forest (IF) - 100 estimators.
\end{itemize}

\subsection{Comparison}
\input{tables/features_studies}

In order to compare the performance of selected feature sets, we reproduced the 16 feature sets marked as ``Feature Reproducible" in Table~\ref{tab:papers}.
Four of the studies \cite{hmm-model, dictionaries, cont-auth-deep-learning,swipe-gesture} implement the exact same group of features as other ones in the set, leaving us with 12 unique and complete feature sets.
More details for each feature set are given in Table~\ref{tab:features_studies}.
Some of the studies enhance their touch-based features with auxiliary data, such as ones coming from the accelerometer or gyroscope, however, we do not reproduce these features due to the lack of such data across all datasets.

We compared the 9 most frequently used classifiers in touch-based authentication studies as shown in Section~\ref{sec:prev_class} across all of the 12 feature sets and report their performance in EER.
We also compared all four aggregation techniques described in Section~\ref{sec:prev_agg} to highlight the best-performing method.
In addition, we include the analysis of the \textit{Median} of scores as an alternative to the \textit{Mean} approach.
The aggregation window we chose in this set of experiments is 5 based on the availability of data and the diminishing returns of larger window sizes as shown in the related literature \cite{common-evaluation,touchalytics}.
For this comparison, we use the novel classifier and feature set described below.

We also introduce 3 novel techniques which we have not identified in other touch-based authentication studies.
We include these in our final analysis:

\subsubsection{Novel feature set} 

We compiled a new feature set (\textit{ALL}) by implementing all \textit{swipe-based} features from our literature review.
These are derived from the X, Y, pressure, and area values of a swipe as described in Appendix~\ref{sec:all_features}.

In addition to this, we utilized a feature selection algorithm which reduces the total number of features from the dataset.
The goal of such approaches is to ensure better computational performance and overall results.
For instance, this can be achieved by pruning features that contribute little to the output of the classifier or even have a negative effect on it.
The feature selection algorithm we use is Analysis of Variance (\textit{ANOVA}) using the F-value between features.
In order to ensure the method generalizes well, we used the three datasets (CEP \cite{common-evaluation}, Bioident \cite{information-revealed} and Touchalytics \cite{touchalytics}).
We first selected n number of features for each dataset using ANOVA.
Then, we only kept features that are sampled in at least two of the three datasets.
We experimented with sizes of 50, 75, 100, and 125 for the parameter n.
In our preliminary results, we established that $n=125$ is the best-performing one in our case and we use it for the rest of this study.
However, we highlight that in this case, the general method for feature selection is more important for further research or industry applications rather than the individual features we chose.

\subsubsection{Novel classifier} 
We propose an ensemble method (\textit{ENS}) for classification based on a combination of results from other classifiers.
Ensemble methods are a well-known strategy used to combine multiple machine learning models which produce a result better than the outcome of each individual classifier.
This is due to the fact that on some examples, some classifiers might perform poorly but on average models will agree on the correct decision.
The algorithm we use outputs a final score by averaging out the probabilities from the predictions of the best-performing individual classifiers.
We performed preliminary experiments with three different combinations of classifiers of sizes 3 (SVM, RF, NN), 5 (SVM, RF, NN, kNN, LR), and 7 (SVM, RF, NN, kNN, LR, NB, DT) and found that the best-performing one in our case is the one consisting of SVM, Random Forest, and Neural Network.
Similar to the novel feature set selection, the specific group of classifiers that we chose is less of interest than the proposed method itself.

\input{tables/feature_class}
\input{tables/aggregations}
    
\subsubsection{Novel aggregation method}
We introduce a Recurrent Neural Network (RNN) stacking algorithm which takes as an input a list of scores and outputs a final probability based on them.
Stacking involves training a model on the outputs of other models in order to produce a final result, much like other aggregation methods described in Section~\ref{sec:prev_agg}.
In addition, we hypothesized that the sequential nature of the RNN processing would work well with the temporal nature of swipes in touch dynamics.
As such our RNN model consists of a single hidden Long Short-Term Memory (LSTM) layer of size 20 with a 'tanh' activation function and one 'sigmoid' output layer. 
We use an LSTM layer in order to avoid the 'vanishing gradient problem' associated with traditional RNNs where long-term gradients tend to go to 0 (vanish) or explode and go to infinity.
Similar to the NN described in Section~\ref{sec:prev_class}, we use the 'Adam' optimizer with a binary cross-entropy as a loss function.

\subsection{Results}
\label{sec:results}

The results from the feature set and classifier comparisons can be found in Table~\ref{tab:feat_class_results}.
On average, the best-performing feature set is the one generated from the \textit{ANOVA} method which we proposed with an average of 17.05\% EER across all classifiers.
Furthermore, using \textit{ALL} identified \textit{swipe-based} features resulted in a similar performance.
The lowest EER from the studies we re-implemented was \cite{towards-continuous-passive} with an average of 17.48\% EER over all the machine learning algorithms examined.
We attribute the low performance of some feature sets such as \cite{explainability} (28.02\%) to the small number of features included.
However, the final model of the study uses additional features from sensor data which could result in much better overall performance.

In terms of classifiers, on average, our ensemble method was the best-performing with an average of 14.51\% EER across all feature sets.
The three individual classifiers in the ensemble (SVM, RF, and NN) also form a well-performing group with 15.67\%, 15.49\%, and 15.20\% respectively.
The one-class classifiers (OC-SVM and IF) produced the worst results in our experiments.
Overall the best-performing model consisted of our proposed \textit{ANOVA} feature set and ensemble classifier with an overall performance of 11.67\% EER for a single swipe.

However, touch-based systems should not operate on a single swipe as it can be insufficient for authentication. 
Aggregation methods, based on multiple consecutive swipes result in improved system performance.
The results from the aggregation experiment can be found in Table~\ref{tab:agg_results}.
The proposed RNN stacking algorithm was the best-performing method in the aggregation comparison with 6.27\% EER.
The \textit{Mean} (6.35\%), \textit{Median} (6.47\%), \textit{Trust} (6.55\%) and \textit{Feed} (7.07\%) methods resulted in very similar performance.
The worst performing aggregation method in our experiments was \textit{Vote} with 10.59\% EER.
Nevertheless, all of the aggregation methods perform better than using a single swipe to make authentication decisions.

\subsubsection{Dataset comparison}

\input{figures/combined}

In order to determine the generality of our results, we replicated our experiments across three datasets - CEP \cite{common-evaluation}, Bioident \cite{information-revealed} and Touchalytics \cite{touchalytics}.
When examining each category (features, classifiers, and aggregations) we choose the best-performing methods determined in our previous experiments and keep them constant (e.g., for all classifiers comparisons, we use the best-performing feature set).

The results of our comparison across all three datasets can be found in Figure~\ref{fig:combined}.
Our ANOVA-based feature set and the ensemble classifier consistently outperform the other methods across all three datasets.
However, the RNN stacking algorithm was the third-best-performing on the Bioident and Touchalytics datasets.
The \textit{Trust} and \textit{Feed} aggregation methods were the best-performing on these datasets by a margin of 0.52\% and 1.78\% EER respectively.

Xu et. al. \cite{towards-continuous-passive} and Frank et al. \cite{touchalytics} are other consistently well-performing feature sets with 37 and 31 features respectively.
Similarly, the SVM, RF, and NN classifiers provide stable performance across all the datasets examined. 

\section{Discussion}
\label{sec:discussion}

This study serves as an overview and improvement over the techniques for feature extraction, classification, and aggregation used in touch-based authentication studies. 
The experiments conducted in Section~\ref{sec:experiments} suggest that when performance is the only concern for a touch-based authentication system, the optimal model would make use of our proposed feature set, ensemble classifier, and RNN stacking aggregation which achieves an EER of 6.28\% using a window of 5 swipes on the large CEP dataset.
However, the lowest EER we achieved using this model is 4.80\% by increasing the aggregation window to 16 strokes.

The relative performance benefits of each individual technique are shown in Table~\ref{tab:results}.
We highlight the next best and the median performing techniques while featuring the difference in EER with the novel ones proposed in this paper.
While the improvements might be perceived as marginal compared to the second-best methods, they are quite significant compared to the median, particularly in the feature extraction and classification cases.
These results highlight the importance of fair comparison between models which can be helpful for decision-making in the broader touch-based authentication community.


It is worth noting that the results in our experiments might not match the results originally reported by a particular study, sometimes by a large margin.
For instance, the EER we obtain using the Touchalytics \cite{touchalytics} feature set on their dataset is multiple times higher than the one attained in the original study.
This is due to the fair evaluation practices we follow as described in Section~\ref{sec:eval_methods}.
Substantially less (16) of the original 41 users in the dataset fit into our criteria and were used in our evaluation.
Many of them had done only 2 sessions, resulting in training and testing data skew closer to 50\%/50\% rather than the target of 80\%/20\%. 
Furthermore, we report the mean EER, while the original study reports the median.

\input{tables/results}

The results we obtained, however, do not guarantee that the techniques we propose are best suited for all datasets and applications.
For instance, our models minimize Equal Error Rates but might require more resources leading to computational power needs and time to execute.
To this end, our analysis examines a large number of techniques and these can be used as alternatives for each stage of the touch-authentication lifecycle.

We have established that the best-performing model in terms of EER consists of our \textit{ANOVA} feature set, ensemble classifier (with SVM, RF, and NN), and a stacking LSTM-based with an EER of 6.28\%. 
However, computational performance could limit the possibility of using this model in practice, particularly due to the mobile environment it is intended for.
In this case, we recommend the use of less sophisticated architecture to be used which can still deliver similar results but at a more cost-efficient computational performance.
Furthermore, we found that there is some variation between the results we obtained on each dataset.
For this reason, selecting consistently well-performing models might be preferred for some applications.
While EER is a good measure for the overall performance of biometric systems, in continuous authentication the focus can be on guaranteeing a low False Negative Rate to ensure adequate usability of the system.
However, that is application-specific and requires further examination which is beyond the scope of this paper.

Even though we ultimately performed the comparison on all three datasets, we believe the results on the CEP dataset are the most representative.
That is due to the larger size in terms of users, sessions performed, and the length of each session.

\subsection{Limitations}

There are several limitations to our experimental approach and results.
Firstly, the implementation details of some features, classifiers and aggregation methods might not be perfectly reproduced from related work, despite our best effort.
Furthermore, the categories we have grouped techniques in might be quite broad with many internal differences between studies.
For instance, implementing a generic trust model algorithm will not necessarily represent the nuances of all models falling under this category.
Similarly, Neural Network implementations may vary between papers and differ from our architecture, and optimizing the hyperparameters of models might lead to better overall performance.
We believe that one-class classifiers in particular can achieve better results by fine-tuning the parameters.

Furthermore, some of the classification algorithms and feature sets that are not as prevalent in the field or are not reproducible might outperform the more popular methods we examine. 
Finally, the fact that the methods we examine are mostly consistent throughout the three datasets is encouraging, however, application to other touch-based authentication datasets might result in much different behavior.

\section{Conclusion}

In this paper, we performed a comprehensive review of the approaches for feature extraction, classification, and aggregation in the field. 
We investigated the prevalence of each technique in the relevant literature and categorized the feature extraction and aggregation methods.
Furthermore, we presented and described a set of 149 unique features extracted from related work and identified 9 publicly available datasets for touch-based authentication.
We benchmarked the performance of the most common feature sets, classifiers, and aggregation methods in the field with a set of experiments consisting of a total of 219 model configurations.
We introduced a novel feature set, ensemble-based classifier, and an RNN-based stacking aggregation method that outperform the state-of-the-art by 0.79\%, 0.69\%, and 0.07\% EER respectively. 
Finally, we concluded that our findings are largely similar across multiple datasets and provided a discussion of our results, including the limitation of the study.

\bibliographystyle{ACM-Reference-Format}
\bibliography{references.bib}



\appendix

\section{Abbreviations}
\label{sec:abbre}

\subsection{Classifiers}
\begin{enumerate}
    \item[] \textbf{AB} - AdaBoost
    \item[] \textbf{BN} - Bayesian Network
    \item[] \textbf{CPANN} - Counter Propagation Artificial Neural Network
    \item[] \textbf{DT} - Decision Tree 
    \item[] \textbf{EE} - Elliptic Envelop
    \item[] \textbf{ENS} - Ensemble
    \item[] \textbf{GB} - Gradient Boosting
    \item[] \textbf{HMM} - Hidden Markov Models
    \item[] \textbf{IF} - Isolation Forest
    \item[] \textbf{KDTGR} - Kernel Dictionary-based Touch Gesture Recognition
    \item[] \textbf{KSRC} - Kernel Sparse Representation-based Classification
    \item[] \textbf{LOF} - Local Outlier Factor
    \item[] \textbf{LR} - Logistic Regression
    \item[] \textbf{NB} - Naive Bayes
    \item[] \textbf{NN} - Neural Networks
    \item[] \textbf{OC-SVM} - OneClass Support Vector Machine
    \item[] \textbf{PSO-RBFN} - Particle Swarm Optimization Radial Basis Function Network
    \item[] \textbf{RC} - Random Committee
    \item[] \textbf{RF} - Random Forest
    \item[] \textbf{SM} - Scaled Manhattan
    \item[] \textbf{SVM} - Support Vector Machine
    \item[] \textbf{StrOUD} - Strangeness based OUtlier Detection
    \item[] \textbf{kNN} - k Nearest Neighbors
\end{enumerate}

\subsection{Metrics}
\begin{enumerate}
    \item[] \textbf{ACC} - Accuracy
    \item[] \textbf{ANGA} - Average Number of Genuine Actions
    \item[] \textbf{ANIA} - Average Number of Impostor Actions
    \item[] \textbf{AUC} - Area Under Curve
    \item[] \textbf{FAR} - False Acceptance Rate
    \item[] \textbf{FRR} - False Rejection Rate
    \item[] \textbf{HTER} - Half Total Error Rate
    \item[] \textbf{ROC} - Receiver Operating Characteristic
\end{enumerate}

\section{All Features}
\label{sec:all_features}
\input{tables/new_features}

\end{document}

%% file: figures/lifecycle.tex
\begin{figure*}[!t]
\centering
\includegraphics[width=1\linewidth]{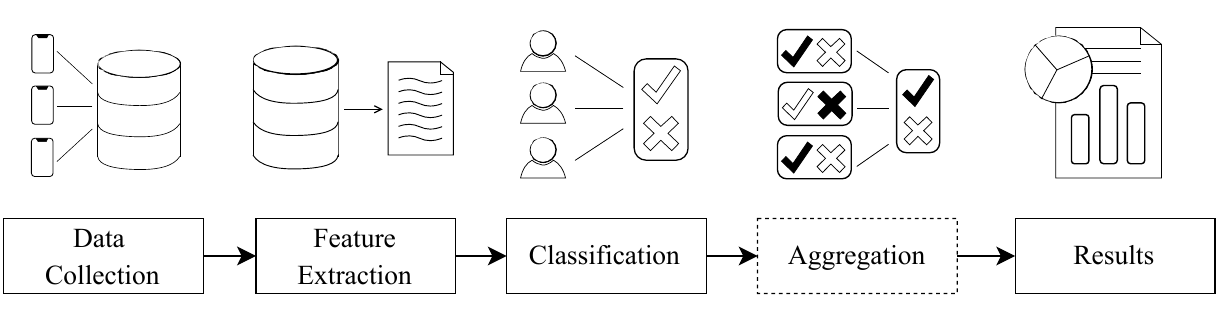}
\caption{Lifecycle of a touch-based authentication system. The aggregation step is optional. The focus of this paper is on the feature extraction, classification, and aggregation methods.}
\vspace{6px}
\label{fig:lifecycle}
\end{figure*}

%% file: tables/datasets.tex
\vspace{10px}
\begin{table*}[!t]
\renewcommand{\arraystretch}{1.2}
\caption{Publicly available touch-dynamics datasets. \reproducible~- currently accessible without additional processes, \partlyreproducible~- can be accessed through email or special process, \unreproducible~- link or instructions currently not working. Links accessed on 20 January 2022. The ``Usable" column denotes the largest group of users with the same phone model, at least two sessions, and data for coordinates, pressure, and area.}
\vspace{6px}
\label{tab:datasets}
\centering
\begin{tabularx}{\textwidth}{p{2.8cm}r>{\raggedleft\arraybackslash}p{1.8cm}>{\raggedleft\arraybackslash}p{1.8cm}>{\raggedleft\arraybackslash}p{1.8cm}>{\raggedleft\arraybackslash}X}
\toprule
Dataset & \makecell[r]{Publication \\ Year} & \makecell[r]{Total Users \\ (Usable)} & Sessions & Accessible & Notes \\
\midrule
Touchalytics \cite{touchalytics} &  2012  & 41 (15) &  3  & \reproducible \footnotemark[1] & - \\
WVW \cite{which-verifiers-work} & 2013  & 190 (0) & 2 & 
 \unreproducible \footnotemark[2] & Data currently not accessible\\
TCPA \cite{towards-continuous-passive} & 2014  & 32 (0) & 1 & 
 \partlyreproducible \footnotemark[3] & \makecell[r]{Participants conduct only a single session}\\
UMDAA-02 \cite{active-user-authentication} & 2016 & 48 (0) & \makecell[r]{11-429} & \reproducible \footnotemark[4] & Touch area values unavailable \\
Bioident \cite{information-revealed} & 2016  & 71 (26) &  1-4  & \reproducible \footnotemark[5] & - \\
TGA \cite{posture-size-config} & 2019  & 31 (0) & 8 & \partlyreproducible \footnotemark[6] & \makecell[r]{Public data contains only extracted features} \\
Brainrun \cite{brainrun} & 2020  & 2344 (0) & 1-1105 & \reproducible \footnotemark[7] & \makecell[r]{Touch area and pressure values unavailable} \\
HuMIdb \cite{humdb-1,humdb-2} & 2020  & 600 (0) & 1-5 & \partlyreproducible \footnotemark[8] & \makecell[r]{Each session contains only a single swipe} \\
CEP \cite{common-evaluation} & 2022  & 470 (64) &  1-30 & \reproducible \footnotemark[9] & - \\
\bottomrule
\end{tabularx}
\end{table*}

%% file: tables/studies.tex
\begin{table*}[!t]
\renewcommand{\arraystretch}{1.2}
\centering
    \caption{Techniques in touch-based authentication studies. Classifier and metric abbreviations are given in Appendix~\ref{sec:abbre}. The following symbols are used in the table: \textbf{?}~- unclear, \reproducible~- we can completely reproduce the features described in the paper and can compare it with other feature sets, \partlyreproducible~- we can reproduce part of the features described in the study but cannot compare it with other feature sets, \unreproducible~- features are not described well enough to be reproduced.}
    \vspace{6px}
	\label{tab:papers}
	\begin{tabularx}{\textwidth}{p{2cm}p{2.9cm}lrrrr}
	\toprule
	\makecell[l]{Study \\ (Year)} & \makecell[l]{Feature Extraction \\ (Count)} & \makecell[l]{Feature \\ Reprod.} & Classifiers & Metrics & \makecell[r]{Aggregation} \\
	\midrule
	\cite{continuous-mobile-touchscreen-gestures}~~~(2012) &  Swipe-based (53) & \unreproducible & DT, RF, BN & FAR, FRR & Vote
	\\
	\cite{touchalytics}~~~(2012) & Swipe-based (31) & \reproducible & kNN, SVM & EER & Mean
	\\
	\cite{unobservable-re-authentication}~~~(2013) & Swipe-based (13) & \reproducible & SVM & ACC & Feed
	\\
	\cite{silent-sense}~~~(2013) & Swipe-based (?) & \unreproducible & OC-SVM, SVM & FAR, FRR, ACC & Trust
	\\
	\cite{which-verifiers-work}~~~(2013) & Swipe-based (28) & \reproducible &  \makecell[r]{LR, SVM, RF, NB, NN, kNN,\\ BN, SM, Euclidian, DT} & EER &  Feed
	\\
	\cite{tips}~~~(2014) & Swipe-based (?) & \unreproducible & kNN & ACC & Feed
	\\
	\cite{latent-gestures}~~~(2014) & Swipe-based (?) & \partlyreproducible & IF, SVM, NB, BN, RF & ACC &  Feed
	\\
	\cite{design-of-touch}~~~(2014) & Session-based (8) & \notapplic & \makecell[r]{DT, NB, RBFN,\\ PSO-RBFN, NN, kNN} & FAR, FRR & Other
	\\
	\cite{hmm-model}~~~(2014) & Swipe-based (31) & \reproducible & HMM & EER, FAR, FRR & Mean
	\\
	\cite{towards-continuous-passive}~~~(2014) & Swipe-based (37) & \reproducible & SVM & ACC, AER &  Mean
	\\
	\cite{statistical-images,graphic-feature}~~~(2014) & Image-based & \notapplic & Proprietary & ACC, EER & Other
	\\
	\cite{performance-analysis}~~~(2015) & Swipe-based (58) & \reproducible & kNN, SVM, NN, RF & ROC, FRR, FAR & Feed
	\\
	\cite{dictionaries}~~~(2015) & Swipe-based (27) & \reproducible & SVM, KSRC, KDTGR & EER & Mean
	\\
	\cite{information-revealed}~~~(2015) & Swipe-based (15) & \reproducible  & kNN, RF, SVM & ACC & Trust
	\\
	\cite{swipe-gesture}~~~(2015) & Swipe-based (15) & \reproducible & NN, CPANN & ANGA, ANIA & Trust
	\\
	\cite{power-consumption}~~~(2015) & Swipe-based (5) & \reproducible & StrOUD & ROC, EER & N/A 
	\\
	\cite{fusing-swiping}~~~(2016) & Session-based (5) & \notapplic & kNN, RF & FAR, FRR, ACC & Other
	\\
	\cite{active-user-authentication}~~~(2016) & Swipe-based (24) & \reproducible & \makecell[r]{kNN, SVM, NB, LR, RF, GB} & EER & Mean 
	\\
	\cite{trace-maps}~~~(2017) & Image-based & \notapplic & SVM, DT, RF, NB & ACC & Other
	\\
	\cite{towards-cont-passive}~~~(2017) & Swipe-based (59) & \partlyreproducible & SVM, RF, DT & AUC & N/A
	\\
	\cite{one-class}~~~(2018) & Session-based (5) & \notapplic & \makecell[r]{AB, NB, kNN, LDA, LR, NN, RF,\\ SVM, OC-SVM, LOF, IF, EE} & \makecell[r]{FAR, FRR, HTER, AUC} & N/A
	\\
	\cite{touch-wb}~~~(2018) & Session-based (21) & \notapplic  & \makecell[r]{DT, NB, Kstar, RBFN, NN, PSO-RBFN} & FAR, FRR, AER & Other
	\\
	\cite{touch-pattern-isolation}~~~(2018) &  Swipe-based (8) & \reproducible  & IF & ANGA,ANIA & Trust
	\\
	\cite{cont-auth-deep-learning}~~~(2019) &  Swipe-based (28) & \reproducible & NN & ACC, EER & N/A
	\\
	\cite{posture-size-config}~~~(2019) & Swipe-based (18) & \reproducible  & RF, SVM, LR, NB, NN & EER & Vote
	\\
	\cite{touch-metric}~~~(2019) & Swipe-based (18) & \partlyreproducible & NB, NN, RC, RF, BN, DT & ACC & N/A
	\\
	\cite{behave-sense}~~~(2019) & Swipe-based (16) & \partlyreproducible & IF, OC-SVM & ACC & Trust
	\\
	\cite{auto-sen}~~~(2020) & Swipe-based (?) & \unreproducible & NN & FAR, FRR, EER & N/A
	\\
	\cite{explainability}~~~(2021) & Swipe-based (12) & \reproducible  & NN & \makecell[r]{ACC, AUC, FRR, FAR}  & Mean
	\\
	\cite{dakota}~~~(2021) & Swipe-based (30) & \reproducible  & \makecell[r]{OC-SVM, kNN, NN, DT, RF, NB} & \makecell[r]{ACC, FAR, FRR, EER, ROC} & N/A
	\end{tabularx}

\end{table*}

%% file: tables/features_studies.tex
\vspace{10px}
\begin{table}[!t]
\renewcommand{\arraystretch}{1.2}
\caption{Reproducible feature sets used in the performance comparison. 
The additional (non-touch-based) features were used in the final proposed model by the paper.
However, we do not re-implement them due to the lack of such data in all of our datasets.}
\vspace{6px}
\label{tab:features_studies}
\centering
\begin{tabular}{lrrr}
\toprule
Study & \makecell[r]{Year of \\  Proposal} & \makecell[r]{Touch \\ Features \\ Count} &  \makecell[r]{Additional \\ Features}\\
\midrule
Frank et al. \cite{touchalytics} &  2013  & 30 & \xmark \\
Li et al. \cite{unobservable-re-authentication} &  2013  & 14 & \xmark \\
Serwadda et al. \cite{which-verifiers-work} &  2013  & 28 & \xmark \\
Xu et al. \cite{towards-continuous-passive} &  2014  & 37 & \xmark \\
Murmuria et al. \cite{power-consumption} &  2015  & 5 & (\textit{sensors, power})~\cmark \\
Antal et al. \cite{information-revealed} &  2015  & 15 & \xmark \\
Mahbub et al. \cite{active-user-authentication} &  2016  & 24 & \xmark \\
Shen et al. \cite{performance-analysis} &  2016  & 58 & \xmark \\
Filippov et al. \cite{touch-pattern-isolation} &  2018  & 11 & \xmark \\
Syed et al. \cite{posture-size-config} &  2019  & 18 & \xmark \\
Rocha et al. \cite{explainability} &  2021  & 12 & \xmark \\
Incel et al. \cite{dakota} &  2021  & 30 & (\textit{sensors})~\cmark \\
\bottomrule
\end{tabular}
\end{table}


%% file: tables/feature_class.tex
\begin{table*}[th]


\begin{center}
\caption{Performance of classifiers applied to different feature sets on the CEP dataset. No aggregation methods are used and the results are reported in EER (\%). The average of each row and column is given.}
\vspace{6px}
\label{tab:feat_class_results}

\begin{tabularx}{\textwidth}{l>{\centering\arraybackslash}X>{\centering\arraybackslash}X>{\centering\arraybackslash}X>{\centering\arraybackslash}X>{\centering\arraybackslash}X>{\centering\arraybackslash}X>{\centering\arraybackslash}X>{\centering\arraybackslash}X>{\centering\arraybackslash}X>{\centering\arraybackslash}X>{\centering\arraybackslash}Xc}

\toprule
Features & \multicolumn{1}{c}{SVM} & \multicolumn{1}{c}{RF} & \multicolumn{1}{c}{NN} & \multicolumn{1}{c}{NB} & \multicolumn{1}{c}{BN} & \multicolumn{1}{c}{KNN} & \multicolumn{1}{c}{DT} & \multicolumn{1}{c}{LR} & \multicolumn{1}{c}{OC-SVM} & \multicolumn{1}{c}{IF} & \multicolumn{1}{c}{ENS} & \multicolumn{1}{c}{~~~~~~~~~~~}\\
\midrule
\cite{touchalytics} & \multicolumn{1}{R}{14.15} & \multicolumn{1}{R}{13.75} & \multicolumn{1}{R}{13.48} & \multicolumn{1}{R}{21.30} & \multicolumn{1}{R}{18.67} & \multicolumn{1}{R}{16.76} & \multicolumn{1}{R}{22.41} & \multicolumn{1}{R}{18.06} & \multicolumn{1}{R}{25.80} & \multicolumn{1}{R}{26.22} & \multicolumn{1}{R}{12.86} & ~18.50 \\
\cite{unobservable-re-authentication} & \multicolumn{1}{R}{15.09} & \multicolumn{1}{R}{14.64} & \multicolumn{1}{R}{14.60} & \multicolumn{1}{R}{21.77} & \multicolumn{1}{R}{18.51} & \multicolumn{1}{R}{17.48} & \multicolumn{1}{R}{23.37} & \multicolumn{1}{R}{20.50} & \multicolumn{1}{R}{24.59} & \multicolumn{1}{R}{26.74} & \multicolumn{1}{R}{13.91} & ~19.20 \\
\cite{which-verifiers-work} & \multicolumn{1}{R}{14.10} & \multicolumn{1}{R}{14.56} & \multicolumn{1}{R}{13.57} & \multicolumn{1}{R}{20.54} & \multicolumn{1}{R}{18.07} & \multicolumn{1}{R}{16.26} & \multicolumn{1}{R}{22.88} & \multicolumn{1}{R}{17.84} & \multicolumn{1}{R}{23.97} & \multicolumn{1}{R}{25.39} & \multicolumn{1}{R}{13.10} & ~18.21 \\
\cite{towards-continuous-passive} & \multicolumn{1}{R}{13.50} & \multicolumn{1}{R}{13.40} & \multicolumn{1}{R}{13.22} & \multicolumn{1}{R}{19.94} & \multicolumn{1}{R}{16.12} & \multicolumn{1}{R}{16.12} & \multicolumn{1}{R}{22.08} & \multicolumn{1}{R}{17.79} & \multicolumn{1}{R}{23.75} & \multicolumn{1}{R}{23.90} & \multicolumn{1}{R}{12.46} & ~17.48 \\
\cite{performance-analysis} & \multicolumn{1}{R}{15.79} & \multicolumn{1}{R}{15.36} & \multicolumn{1}{R}{14.97} & \multicolumn{1}{R}{23.17} & \multicolumn{1}{R}{20.46} & \multicolumn{1}{R}{19.23} & \multicolumn{1}{R}{24.00} & \multicolumn{1}{R}{19.01} & \multicolumn{1}{R}{27.46} & \multicolumn{1}{R}{29.19} & \multicolumn{1}{R}{14.36} & ~20.27 \\
\cite{information-revealed} & \multicolumn{1}{R}{14.00} & \multicolumn{1}{R}{14.39} & \multicolumn{1}{R}{13.95} & \multicolumn{1}{R}{20.79} & \multicolumn{1}{R}{18.30} & \multicolumn{1}{R}{16.34} & \multicolumn{1}{R}{22.88} & \multicolumn{1}{R}{19.02} & \multicolumn{1}{R}{23.66} & \multicolumn{1}{R}{24.41} & \multicolumn{1}{R}{13.32} & ~18.28 \\
\cite{power-consumption} & \multicolumn{1}{R}{20.43} & \multicolumn{1}{R}{18.78} & \multicolumn{1}{R}{19.60} & \multicolumn{1}{R}{24.45} & \multicolumn{1}{R}{22.16} & \multicolumn{1}{R}{21.70} & \multicolumn{1}{R}{25.62} & \multicolumn{1}{R}{25.41} & \multicolumn{1}{R}{26.58} & \multicolumn{1}{R}{26.13} & \multicolumn{1}{R}{18.62} & ~22.68 \\
\cite{active-user-authentication} & \multicolumn{1}{R}{15.83} & \multicolumn{1}{R}{15.69} & \multicolumn{1}{R}{15.28} & \multicolumn{1}{R}{22.55} & \multicolumn{1}{R}{19.72} & \multicolumn{1}{R}{18.07} & \multicolumn{1}{R}{24.23} & \multicolumn{1}{R}{20.43} & \multicolumn{1}{R}{27.27} & \multicolumn{1}{R}{27.45} & \multicolumn{1}{R}{14.60} & ~20.10 \\
\cite{touch-pattern-isolation} & \multicolumn{1}{R}{15.17} & \multicolumn{1}{R}{15.88} & \multicolumn{1}{R}{15.22} & \multicolumn{1}{R}{21.15} & \multicolumn{1}{R}{19.71} & \multicolumn{1}{R}{16.86} & \multicolumn{1}{R}{23.78} & \multicolumn{1}{R}{21.51} & \multicolumn{1}{R}{23.17} & \multicolumn{1}{R}{24.29} & \multicolumn{1}{R}{14.67} & ~19.22 \\
\cite{posture-size-config} & \multicolumn{1}{R}{16.71} & \multicolumn{1}{R}{15.71} & \multicolumn{1}{R}{16.00} & \multicolumn{1}{R}{23.27} & \multicolumn{1}{R}{19.07} & \multicolumn{1}{R}{18.99} & \multicolumn{1}{R}{23.83} & \multicolumn{1}{R}{21.78} & \multicolumn{1}{R}{27.68} & \multicolumn{1}{R}{27.09} & \multicolumn{1}{R}{15.13} & ~20.48 \\
\cite{explainability} & \multicolumn{1}{R}{25.54} & \multicolumn{1}{R}{24.74} & \multicolumn{1}{R}{24.70} & \multicolumn{1}{R}{29.99} & \multicolumn{1}{R}{26.39} & \multicolumn{1}{R}{26.50} & \multicolumn{1}{R}{31.07} & \multicolumn{1}{R}{28.69} & \multicolumn{1}{R}{33.06} & \multicolumn{1}{R}{33.48} & \multicolumn{1}{R}{24.02} & ~28.02 \\
\cite{dakota} & \multicolumn{1}{R}{13.68} & \multicolumn{1}{R}{14.25} & \multicolumn{1}{R}{13.32} & \multicolumn{1}{R}{21.19} & \multicolumn{1}{R}{19.99} & \multicolumn{1}{R}{16.52} & \multicolumn{1}{R}{22.72} & \multicolumn{1}{R}{17.71} & \multicolumn{1}{R}{24.31} & \multicolumn{1}{R}{25.62} & \multicolumn{1}{R}{12.79} & ~18.37 \\
All & \multicolumn{1}{R}{12.77} & \multicolumn{1}{R}{12.78} & \multicolumn{1}{R}{12.45} & \multicolumn{1}{R}{21.67} & \multicolumn{1}{R}{16.21} & \multicolumn{1}{R}{15.86} & \multicolumn{1}{R}{21.71} & \multicolumn{1}{R}{15.88} & \multicolumn{1}{R}{23.93} & \multicolumn{1}{R}{25.03} & \multicolumn{1}{R}{11.70} & ~17.27 \\
ANOVA & \multicolumn{1}{R}{12.57} & \multicolumn{1}{R}{13.00} & \multicolumn{1}{R}{12.36} & \multicolumn{1}{R}{19.84} & \multicolumn{1}{R}{16.37} & \multicolumn{1}{R}{15.63} & \multicolumn{1}{R}{21.94} & \multicolumn{1}{R}{15.81} & \multicolumn{1}{R}{23.45} & \multicolumn{1}{R}{24.96} & \multicolumn{1}{R}{11.67} & ~17.05 \\
\rule{0pt}{3ex} 
& 15.67 & 15.49 & 15.20 & 22.26 & 19.27 & 18.02 & 23.75 & 19.96 & 25.62 & 26.42 & 14.51  \\

\end{tabularx}
\end{center}
\end{table*}

%% file: tables/aggregations.tex
\begin{table*}[th]


\begin{center}
\caption{Performance of aggregation methods on the CEP dataset with \textit{ANOVA} features and an ensable classifier consisting of an SVM, Random Forest and a Neural Network). The aggregation window used is 5 and the results are reported in EER (\%).}
\vspace{6px}
\label{tab:agg_results}

\begin{tabularx}{\textwidth}{>{\centering\arraybackslash}X>{\centering\arraybackslash}X>{\centering\arraybackslash}X>{\centering\arraybackslash}X>{\centering\arraybackslash}X>{\centering\arraybackslash}X}

\toprule
\multicolumn{1}{>{\centering}p{2.62cm}}{Mean} &
\multicolumn{1}{>{\centering}p{2.62cm}}{Median} &
\multicolumn{1}{>{\centering}p{2.62cm}}{Vote} & \multicolumn{1}{>{\centering}p{2.62cm}}{Feed} & \multicolumn{1}{>{\centering}p{2.62cm}}{Trust} & \multicolumn{1}{>{\centering}p{2.62cm}}{Stacking} \\
\midrule
\multicolumn{1}{G}{6.35} &  \multicolumn{1}{G}{6.47} & \multicolumn{1}{G}{10.59} & \multicolumn{1}{G}{7.07} & \multicolumn{1}{G}{6.55} & \multicolumn{1}{G}{6.28} \\
\end{tabularx}
\end{center}
\end{table*}

%% file: figures/combined.tex
\begin{figure*}[!t]
\includegraphics[width=1.00\linewidth]{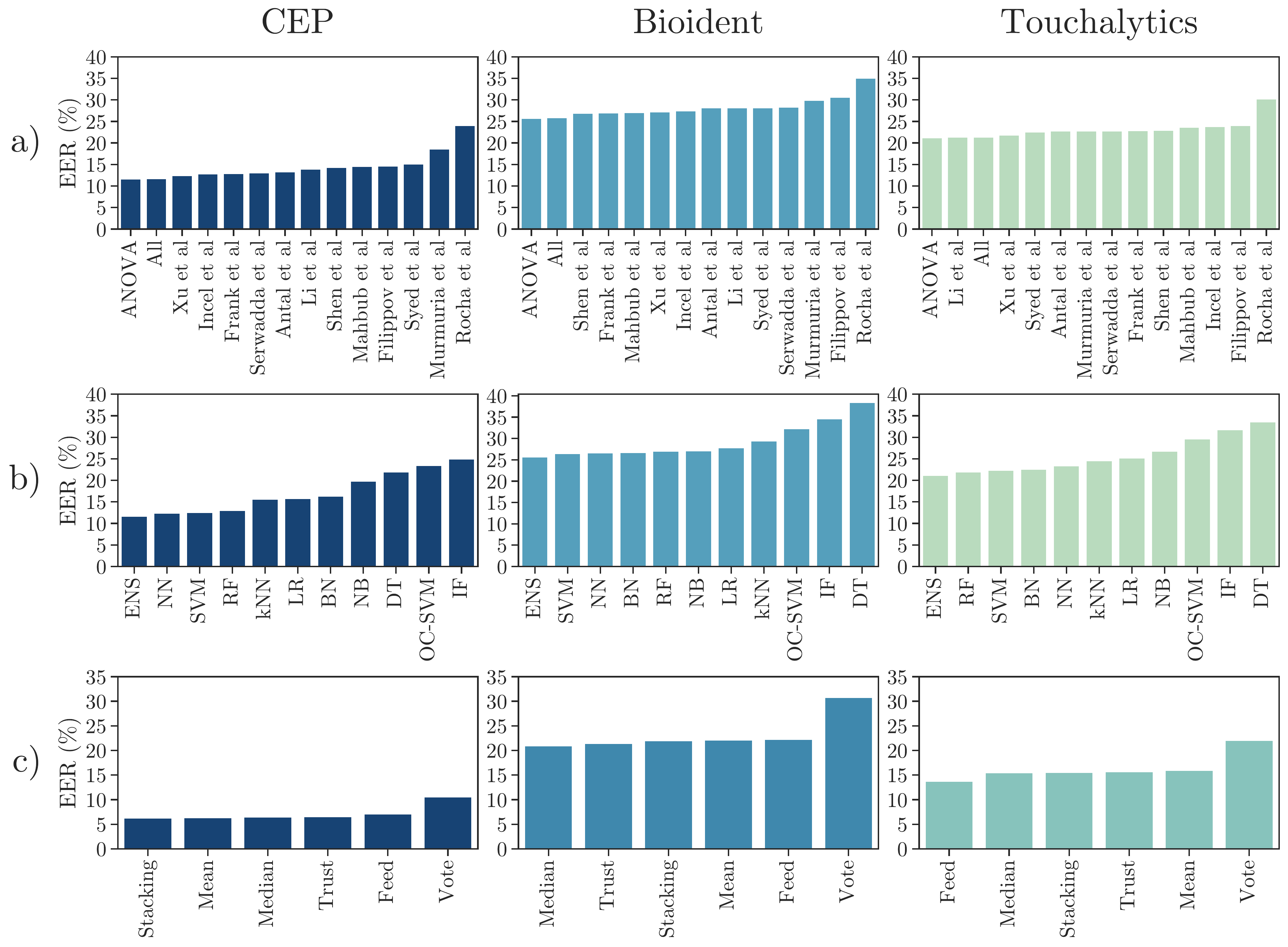}
\caption{Performance of (a) featurests, (b) classifiers, and (c) aggregation methods across CEP, Bioident and Touchalytics touch-based authentication datasets. 
}
\vspace{6px}
\label{fig:combined}
\end{figure*}

%% file: tables/results.tex
\begin{table*}[t]

\renewcommand{\arraystretch}{1.2}
\caption{Difference between the novel techniques proposed in this paper and the next best and median methods in touch-based authentication literature. The differences in EER are reported in percentage points (\%). The best performing feature sets, classifiers and aggregation methods are \textit{ANOVA}, \textit{Ensamble} and \textit{Stacking}.}
\vspace{6px}
\label{tab:results}
\centering
\begin{tabularx}{\textwidth}{l>{\centering\arraybackslash}X>{\centering\arraybackslash}X|>{\centering\arraybackslash}X>{\centering\arraybackslash}X|>{\centering\arraybackslash}X>{\centering\arraybackslash}X}
\toprule
 & \multicolumn{2}{c}{CEP} & \multicolumn{2}{c}{Bioident} & \multicolumn{2}{c}{Touchalytics}  \\
 \cmidrule{2-7} 
 & Next Best & Median & Next Best & Median & Next Best & Median \\
\midrule
Features & \textit{Xu et al.} (+0.79) & +1.94 & \textit{Shen et al.} (+1.22) & +2.08 & \textit{Li et al.} (+0.21) & +1.62 \\
Classification &  NN (+0.69) & +4.14 & SVM (+0.76) & +1.44 & RF (+0.79) & +3.71 \\  
Aggregation  &  \textit{Mean} (+0.07) & +0.27 & \textit{Median} (-1.03) & +0.08 & \textit{Feed} (-1.78) & +0.06 \\
\bottomrule
\end{tabularx}
\end{table*}

%% file: tables/new_features.tex
\begin{table*}[t]
\caption{Geometric features found in related work. Perc. stands for pecentile and Std. Dev. for standard deviation. Full details about each of the features can be found in the corresponding papers. Note that \cite{hmm-model, dictionaries, cont-auth-deep-learning} use the same features as \cite{touchalytics} and \cite{swipe-gesture} uses the same as \cite{information-revealed} except they omit the mid-stroke pressure.} 
\label{tab:features}
\begin{tabularx}{\textwidth}{p{5.3cm}l|p{7cm}l}
\hline
Feature & Studies & Feature & Studies \\ \hline
1-2. Start X,Y & \makecell[l]{\cite{touchalytics, unobservable-re-authentication, towards-continuous-passive, performance-analysis,information-revealed}\\ \cite{active-user-authentication, posture-size-config, dakota, touch-pattern-isolation}} & 43. Std. Dev. acceleration & \cite{which-verifiers-work, performance-analysis} \\
3-4. Stop X,Y & \makecell[l]{\cite{touchalytics, towards-continuous-passive, performance-analysis, information-revealed, active-user-authentication}\\ \cite{posture-size-config, dakota, touch-pattern-isolation}} & \makecell[l]{44-47. First Quartile pressure,\\ area, velocity, acceleration} & \cite{which-verifiers-work} \\
5. Stroke duration & \makecell[l]{\cite{touchalytics, unobservable-re-authentication, which-verifiers-work, towards-continuous-passive, performance-analysis, information-revealed}\\ \cite{power-consumption, active-user-authentication, posture-size-config, dakota, touch-pattern-isolation}} & \makecell[l]{48-51. Third Quartile pressure,\\ area, velocity, acceleration} & \cite{which-verifiers-work} \\
6. End-to-end distance & \makecell[l]{\cite{touchalytics, which-verifiers-work,  towards-continuous-passive, information-revealed, power-consumption}\\ \cite{active-user-authentication, posture-size-config, dakota, touch-pattern-isolation}} & 52-55. Extreme point 1,2 - X,Y & \cite{which-verifiers-work} \\
7. Mid-stroke pressure & \cite{touchalytics, which-verifiers-work,  performance-analysis, information-revealed, posture-size-config, dakota} & 56. Last 2 points tangent & \cite{which-verifiers-work} \\
8. Mid-stroke area & \cite{touchalytics, which-verifiers-work, information-revealed, dakota} & 57. Velocity at first point & \cite{towards-continuous-passive} \\
9. Length of Trajectory & \makecell[l]{\cite{touchalytics, unobservable-re-authentication, which-verifiers-work, towards-continuous-passive, performance-analysis}\\\cite{information-revealed, posture-size-config, dakota, touch-pattern-isolation}} & 58-60. Area, Pressure, Velocity at last point & \cite{towards-continuous-passive} \\
10. Inter-stroke time & \cite{touchalytics, active-user-authentication, posture-size-config} & 61. Last moving direction & \cite{towards-continuous-passive} \\
11. Mean Resultant Length & \cite{touchalytics, information-revealed, active-user-authentication, dakota} & 62. Average points distance & \cite{towards-continuous-passive, performance-analysis, explainability} \\
12. Median acceleration at first 5 points & \cite{touchalytics, active-user-authentication, dakota} & 63. Std. Dev. points distance & \cite{towards-continuous-passive, performance-analysis} \\
13. Median velocity at last 3 points & \cite{touchalytics, active-user-authentication, dakota} & 64-68. LDP X, Y, Area, Pressure, Velocity & \cite{towards-continuous-passive,behave-sense} \\
14. Average velocity & \makecell[l]{\cite{touchalytics, which-verifiers-work,  towards-continuous-passive, performance-analysis}\\\cite{information-revealed, posture-size-config, dakota, touch-pattern-isolation}} & 69-71. Start to LDP Latency, Length, Direction & \cite{towards-continuous-passive} \\
15. Up/Down/Left/Right & \cite{touchalytics, information-revealed, active-user-authentication} & 72-74. LDP to Stop Latency, Length, Direction & \cite{towards-continuous-passive} \\
16. Direction of direct line & \cite{touchalytics, towards-continuous-passive, information-revealed, power-consumption, posture-size-config, dakota} & 75. Ratio distance to LDP Length & \cite{towards-continuous-passive} \\
17. Average direction & \cite{touchalytics} & 76. Total displacement length & \cite{performance-analysis} \\
\makecell[l]{18. Ratio of direct distance\\ to trajectory length} & \cite{touchalytics,  towards-continuous-passive, active-user-authentication, posture-size-config, dakota} & 77. Ratio of displacement and trajectory length & \cite{performance-analysis} \\
19. 20\% perc. velocity & \cite{touchalytics, active-user-authentication, posture-size-config, dakota} & 78-81. Median, IQR, Skewnsess, Kurtosis of distance & \cite{performance-analysis} \\
20. 50\% perc. velocity & \cite{touchalytics, which-verifiers-work,  performance-analysis, active-user-authentication, posture-size-config, dakota} & \makecell[l]{82-86. Avg, Std. Dev, IQR, Skewness,\\ Kurtosis of deviation} & \cite{performance-analysis} \\
21. 80\% perc. velocity & \cite{touchalytics, active-user-authentication, posture-size-config, dakota} & \makecell[l]{87-92. Avg, Median, Std Dev, IQR, Skewness,\\ Kurtosis of pairwise angles} & \cite{performance-analysis} \\
22. 20\% perc. acceleration & \cite{touchalytics, active-user-authentication, dakota} & \makecell[l]{93-98. Avg, Median, Std. Dev, IQR,\\ Skewness, Kurtosis of phase-angles} & \cite{performance-analysis} \\
23. 50\% perc. acceleration & \cite{touchalytics, which-verifiers-work,  performance-analysis, active-user-authentication, dakota} & 99. Displacement to duration ratio & \cite{performance-analysis} \\
24. 80\% perc. acceleration & \cite{touchalytics, active-user-authentication, dakota} & 100-102. IQR, Skewness, Kurtosis of velocities & \cite{performance-analysis} \\
25. 20\% perc. deviation & \cite{touchalytics, active-user-authentication} & \makecell[l]{103-108. Avg, Median, Std. Dev., IQR, Skewness,\\ Kurtosis of angular-velocities} & \cite{performance-analysis} \\
26. 50\% perc. deviation & \cite{touchalytics,  performance-analysis, active-user-authentication} & 109-111. IQR, Skewness, Kurtosis of accelerations & \cite{performance-analysis} \\
27. 80\% perc. deviation & \cite{touchalytics, active-user-authentication} & 112-114. IQR, Skewness, Kurtosis of pressures & \cite{performance-analysis} \\
28. Largest deviation & \cite{touchalytics, information-revealed, active-user-authentication} & 115-116. Min, Max pressure & \cite{towards-cont-passive, explainability} \\
29. Pressure at first point & \cite{unobservable-re-authentication, towards-continuous-passive} & 117-118. Min, Max area & \cite{towards-cont-passive, explainability} \\
30. Area at first point & \cite{unobservable-re-authentication, towards-continuous-passive} & 119-120. Min, Max velocity & \cite{towards-cont-passive, behave-sense} \\
31. First moving direction & \cite{unobservable-re-authentication, towards-continuous-passive} & 121-124. Min, Max, Mean, Median of pressure changes & \cite{towards-cont-passive} \\
32. Average moving direction & \cite{unobservable-re-authentication, information-revealed, active-user-authentication, dakota} & 125-128. Min, Max, Mean, Median of area changes & \cite{towards-cont-passive} \\
33. Average moving curvature & \cite{unobservable-re-authentication} & 129-130. X, Y at max velocity & \cite{behave-sense} \\
34. Average curvature distance & \cite{unobservable-re-authentication} & 131-132. X, Y at min velocity & \cite{behave-sense} \\
35. Average pressure & \cite{unobservable-re-authentication, which-verifiers-work,towards-continuous-passive, performance-analysis, power-consumption} & 133-135. Quadratic fit pressure x2, x, n & \cite{explainability} \\
36. Average touch area & \cite{unobservable-re-authentication, which-verifiers-work, towards-continuous-passive,power-consumption, explainability, touch-pattern-isolation} & 136-138. Min, Max, Avg time duration between points & \cite{explainability} \\
37. Max-area portion & \cite{unobservable-re-authentication} & 139-140. Max deviation of mean X, Y & \cite{dakota} \\
38. Min-pressure portion & \cite{unobservable-re-authentication} & 141-142. 20\% perc. deviation of mean X, Y & \cite{dakota} \\
39. Average acceleration & \cite{which-verifiers-work, performance-analysis} & 143-144. Median deviation of mean X, Y & \cite{dakota} \\
40. Std. Dev. pressure & \cite{which-verifiers-work, towards-continuous-passive, performance-analysis} & 145-146. 80\% perc. deviation of mean X, Y & \cite{dakota} \\
41. Std. Dev. area & \cite{which-verifiers-work, towards-continuous-passive} & 147-148. Direction vector X, Y & \cite{touch-pattern-isolation} \\
42. Std. Dev. velocity & \cite{which-verifiers-work, performance-analysis, active-user-authentication} & 149. Horizontal/Vertical flag & \cite{posture-size-config}
\end{tabularx}
\end{table*}

%% file: main.bbl

\begin{thebibliography}{65}


\ifx \showCODEN    \undefined \def \showCODEN     #1{\unskip}     \fi
\ifx \showDOI      \undefined \def \showDOI       #1{#1}\fi
\ifx \showISBNx    \undefined \def \showISBNx     #1{\unskip}     \fi
\ifx \showISBNxiii \undefined \def \showISBNxiii  #1{\unskip}     \fi
\ifx \showISSN     \undefined \def \showISSN      #1{\unskip}     \fi
\ifx \showLCCN     \undefined \def \showLCCN      #1{\unskip}     \fi
\ifx \shownote     \undefined \def \shownote      #1{#1}          \fi
\ifx \showarticletitle \undefined \def \showarticletitle #1{#1}   \fi
\ifx \showURL      \undefined \def \showURL       {\relax}        \fi
\providecommand\bibfield[2]{#2}
\providecommand\bibinfo[2]{#2}
\providecommand\natexlab[1]{#1}
\providecommand\showeprint[2][]{arXiv:#2}

\bibitem[\protect\citeauthoryear{??}{eri}{[n.d.]}]%
        {ericsson}
 \bibinfo{year}{[n.d.]}\natexlab{}.
\newblock \bibinfo{title}{Ericsson Mobility Report 2021}.
\newblock
  \bibinfo{howpublished}{\url{https://www.ericsson.com/en/reports-and-papers/mobility-report/reports/november-2021}}.
\newblock
\newblock
\shownote{Accessed: 20 January 2022.}


\bibitem[\protect\citeauthoryear{Abadi, Agarwal, Barham, Brevdo, Chen, Citro,
  Corrado, Davis, et~al\mbox{.}}{Abadi et~al\mbox{.}}{2015}]%
        {tensorflow}
\bibfield{author}{\bibinfo{person}{Mart\'{i}n Abadi}, \bibinfo{person}{Ashish
  Agarwal}, \bibinfo{person}{Paul Barham}, \bibinfo{person}{Eugene Brevdo},
  \bibinfo{person}{Zhifeng Chen}, \bibinfo{person}{Craig Citro},
  \bibinfo{person}{Greg~S. Corrado}, \bibinfo{person}{Andy Davis},
  {et~al\mbox{.}}} \bibinfo{year}{2015}\natexlab{}.
\newblock \bibinfo{title}{{TensorFlow}: Large-Scale Machine Learning on
  Heterogeneous Systems}.
\newblock
\newblock
\urldef\tempurl%
\url{https://www.tensorflow.org/}
\showURL{%
\tempurl}
\newblock
\shownote{Software available from tensorflow.org.}


\bibitem[\protect\citeauthoryear{Abuhamad, Abuhmed, Mohaisen, and
  Nyang}{Abuhamad et~al\mbox{.}}{2020}]%
        {auto-sen}
\bibfield{author}{\bibinfo{person}{Mohammed Abuhamad}, \bibinfo{person}{Tamer
  Abuhmed}, \bibinfo{person}{David Mohaisen}, {and} \bibinfo{person}{DaeHun
  Nyang}.} \bibinfo{year}{2020}\natexlab{}.
\newblock \showarticletitle{AUToSen: Deep-Learning-Based Implicit Continuous
  Authentication Using Smartphone Sensors}.
\newblock \bibinfo{journal}{\emph{IEEE Internet of Things Journal}}
  \bibinfo{volume}{7}, \bibinfo{number}{6} (\bibinfo{year}{2020}),
  \bibinfo{pages}{5008--5020}.
\newblock
\urldef\tempurl%
\url{https://doi.org/10.1109/JIOT.2020.2975779}
\showDOI{\tempurl}


\bibitem[\protect\citeauthoryear{Acien, Morales, Fi{\'{e}}rrez,
  Vera{-}Rodr{\'{\i}}guez, and Bartolome}{Acien et~al\mbox{.}}{2020a}]%
        {humdb-1}
\bibfield{author}{\bibinfo{person}{Alejandro Acien}, \bibinfo{person}{Aythami
  Morales}, \bibinfo{person}{Julian Fi{\'{e}}rrez},
  \bibinfo{person}{Rub{\'{e}}n Vera{-}Rodr{\'{\i}}guez}, {and}
  \bibinfo{person}{Ivan Bartolome}.} \bibinfo{year}{2020}\natexlab{a}.
\newblock \showarticletitle{Be-CAPTCHA: Detecting Human Behavior in Smartphone
  Interaction using Multiple Inbuilt Sensors}.
\newblock \bibinfo{journal}{\emph{CoRR}}  \bibinfo{volume}{abs/2002.00918}
  (\bibinfo{year}{2020}).
\newblock
\showeprint[arxiv]{2002.00918}
\urldef\tempurl%
\url{https://arxiv.org/abs/2002.00918}
\showURL{%
\tempurl}


\bibitem[\protect\citeauthoryear{Acien, Morales, Fi{\'{e}}rrez,
  Vera{-}Rodr{\'{\i}}guez, and Delgado{-}Mohatar}{Acien et~al\mbox{.}}{2020b}]%
        {humdb-2}
\bibfield{author}{\bibinfo{person}{Alejandro Acien}, \bibinfo{person}{Aythami
  Morales}, \bibinfo{person}{Julian Fi{\'{e}}rrez},
  \bibinfo{person}{Rub{\'{e}}n Vera{-}Rodr{\'{\i}}guez}, {and}
  \bibinfo{person}{Oscar Delgado{-}Mohatar}.} \bibinfo{year}{2020}\natexlab{b}.
\newblock \showarticletitle{BeCAPTCHA: Bot Detection in Smartphone Interaction
  using Touchscreen Biometrics and Mobile Sensors}.
\newblock \bibinfo{journal}{\emph{CoRR}}  \bibinfo{volume}{abs/2005.13655}
  (\bibinfo{year}{2020}).
\newblock
\showeprint[arxiv]{2005.13655}
\urldef\tempurl%
\url{https://arxiv.org/abs/2005.13655}
\showURL{%
\tempurl}


\bibitem[\protect\citeauthoryear{Ahmad, Sajjad, Jan, Mehmood, Rho, and
  Baik}{Ahmad et~al\mbox{.}}{2017}]%
        {trace-maps}
\bibfield{author}{\bibinfo{person}{Jamil Ahmad}, \bibinfo{person}{Muhammad
  Sajjad}, \bibinfo{person}{Zahoor Jan}, \bibinfo{person}{Irfan Mehmood},
  \bibinfo{person}{Seungmin Rho}, {and} \bibinfo{person}{Sung~Wook Baik}.}
  \bibinfo{year}{2017}\natexlab{}.
\newblock \showarticletitle{Analysis of interaction trace maps for active
  authentication on smart devices}.
\newblock \bibinfo{journal}{\emph{Multimedia Tools and Applications}}
  \bibinfo{volume}{76}, \bibinfo{number}{3} (\bibinfo{date}{01 Feb}
  \bibinfo{year}{2017}), \bibinfo{pages}{4069--4087}.
\newblock
\showISSN{1573-7721}
\urldef\tempurl%
\url{https://doi.org/10.1007/s11042-016-3450-y}
\showDOI{\tempurl}


\bibitem[\protect\citeauthoryear{Antal, Bokor, and Szabó}{Antal
  et~al\mbox{.}}{2015}]%
        {information-revealed}
\bibfield{author}{\bibinfo{person}{Margit Antal}, \bibinfo{person}{Zsolt
  Bokor}, {and} \bibinfo{person}{László~Zsolt Szabó}.}
  \bibinfo{year}{2015}\natexlab{}.
\newblock \showarticletitle{Information revealed from scrolling interactions on
  mobile devices}.
\newblock \bibinfo{journal}{\emph{Pattern Recognition Letters}}
  \bibinfo{volume}{56} (\bibinfo{year}{2015}), \bibinfo{pages}{7--13}.
\newblock
\showISSN{0167-8655}
\urldef\tempurl%
\url{https://doi.org/10.1016/j.patrec.2015.01.011}
\showDOI{\tempurl}


\bibitem[\protect\citeauthoryear{Antal and Szabó}{Antal and Szabó}{2016}]%
        {benchmark-touch}
\bibfield{author}{\bibinfo{person}{Margit Antal} {and}
  \bibinfo{person}{László~Zsolt Szabó}.} \bibinfo{year}{2016}\natexlab{}.
\newblock \showarticletitle{Biometric Authentication Based on Touchscreen Swipe
  Patterns}.
\newblock \bibinfo{journal}{\emph{Procedia Technology}}  \bibinfo{volume}{22}
  (\bibinfo{year}{2016}), \bibinfo{pages}{862 -- 869}.
\newblock
\showISSN{2212-0173}
\urldef\tempurl%
\url{https://doi.org/10.1016/j.protcy.2016.01.061}
\showDOI{\tempurl}
\newblock
\shownote{9th International Conference Interdisciplinarity in Engineering,
  INTER-ENG 2015, 8-9 October 2015, Tirgu Mures, Romania.}


\bibitem[\protect\citeauthoryear{Aviv, Gibson, Mossop, Blaze, and Smith}{Aviv
  et~al\mbox{.}}{2010}]%
        {smudge-attack}
\bibfield{author}{\bibinfo{person}{Adam~J. Aviv}, \bibinfo{person}{Katherine
  Gibson}, \bibinfo{person}{Evan Mossop}, \bibinfo{person}{Matt Blaze}, {and}
  \bibinfo{person}{Jonathan~M. Smith}.} \bibinfo{year}{2010}\natexlab{}.
\newblock \showarticletitle{Smudge Attacks on Smartphone Touch Screens}. In
  \bibinfo{booktitle}{\emph{Proceedings of the 4th USENIX Conference on
  Offensive Technologies}} (Washington, DC) \emph{(\bibinfo{series}{WOOT'10})}.
  \bibinfo{publisher}{USENIX Association}, \bibinfo{address}{USA},
  \bibinfo{pages}{1–7}.
\newblock


\bibitem[\protect\citeauthoryear{Bo, Zhang, Li, Huang, and Wang}{Bo
  et~al\mbox{.}}{2013}]%
        {silent-sense}
\bibfield{author}{\bibinfo{person}{Cheng Bo}, \bibinfo{person}{Lan Zhang},
  \bibinfo{person}{Xiang-Yang Li}, \bibinfo{person}{Qiuyuan Huang}, {and}
  \bibinfo{person}{Yu Wang}.} \bibinfo{year}{2013}\natexlab{}.
\newblock \showarticletitle{SilentSense: Silent User Identification via Touch
  and Movement Behavioral Biometrics}. In \bibinfo{booktitle}{\emph{Proceedings
  of the 19th Annual International Conference on Mobile Computing \&
  Networking}} (Miami, Florida, USA) \emph{(\bibinfo{series}{MobiCom '13})}.
  \bibinfo{publisher}{Association for Computing Machinery},
  \bibinfo{address}{New York, NY, USA}, \bibinfo{pages}{187–190}.
\newblock
\showISBNx{9781450319997}
\urldef\tempurl%
\url{https://doi.org/10.1145/2500423.2504572}
\showDOI{\tempurl}


\bibitem[\protect\citeauthoryear{Buitinck, Louppe, Blondel, Pedregosa, Mueller,
  Grisel, Niculae, Prettenhofer, Gramfort, Grobler, Layton, VanderPlas, Joly,
  Holt, and Varoquaux}{Buitinck et~al\mbox{.}}{2013}]%
        {sklearn_api}
\bibfield{author}{\bibinfo{person}{Lars Buitinck}, \bibinfo{person}{Gilles
  Louppe}, \bibinfo{person}{Mathieu Blondel}, \bibinfo{person}{Fabian
  Pedregosa}, \bibinfo{person}{Andreas Mueller}, \bibinfo{person}{Olivier
  Grisel}, \bibinfo{person}{Vlad Niculae}, \bibinfo{person}{Peter
  Prettenhofer}, \bibinfo{person}{Alexandre Gramfort}, \bibinfo{person}{Jaques
  Grobler}, \bibinfo{person}{Robert Layton}, \bibinfo{person}{Jake VanderPlas},
  \bibinfo{person}{Arnaud Joly}, \bibinfo{person}{Brian Holt}, {and}
  \bibinfo{person}{Ga{\"{e}}l Varoquaux}.} \bibinfo{year}{2013}\natexlab{}.
\newblock \showarticletitle{{API} design for machine learning software:
  experiences from the scikit-learn project}. In \bibinfo{booktitle}{\emph{ECML
  PKDD Workshop: Languages for Data Mining and Machine Learning}}.
  \bibinfo{pages}{108--122}.
\newblock


\bibitem[\protect\citeauthoryear{Cheon, Shin, Huh, Kim, and Oakley}{Cheon
  et~al\mbox{.}}{2020}]%
        {gestures-mturk}
\bibfield{author}{\bibinfo{person}{E. Cheon}, \bibinfo{person}{Y. Shin},
  \bibinfo{person}{J. Huh}, \bibinfo{person}{H. Kim}, {and} \bibinfo{person}{I.
  Oakley}.} \bibinfo{year}{2020}\natexlab{}.
\newblock \showarticletitle{Gesture Authentication for Smartphones: Evaluation
  of Gesture Password Selection Policies}. In \bibinfo{booktitle}{\emph{2020
  IEEE Symposium on Security and Privacy (SP)}}. \bibinfo{publisher}{IEEE
  Computer Society}, \bibinfo{address}{Los Alamitos, CA, USA},
  \bibinfo{pages}{249--267}.
\newblock
\urldef\tempurl%
\url{https://doi.org/10.1109/SP40000.2020.00034}
\showDOI{\tempurl}


\bibitem[\protect\citeauthoryear{Chollet et~al\mbox{.}}{Chollet
  et~al\mbox{.}}{2015}]%
        {keras}
\bibfield{author}{\bibinfo{person}{Fran\c{c}ois Chollet} {et~al\mbox{.}}}
  \bibinfo{year}{2015}\natexlab{}.
\newblock \bibinfo{title}{Keras}.
\newblock \bibinfo{howpublished}{\url{https://keras.io}}.
\newblock


\bibitem[\protect\citeauthoryear{Eberz, Rasmussen, Lenders, and
  Martinovic}{Eberz et~al\mbox{.}}{2017}]%
        {eberz}
\bibfield{author}{\bibinfo{person}{Simon Eberz}, \bibinfo{person}{Kasper~B.
  Rasmussen}, \bibinfo{person}{Vincent Lenders}, {and} \bibinfo{person}{Ivan
  Martinovic}.} \bibinfo{year}{2017}\natexlab{}.
\newblock \showarticletitle{Evaluating Behavioral Biometrics for Continuous
  Authentication: Challenges and Metrics}. In
  \bibinfo{booktitle}{\emph{Proceedings of the 2017 ACM on Asia Conference on
  Computer and Communications Security}} (Abu Dhabi, United Arab Emirates)
  \emph{(\bibinfo{series}{ASIA CCS '17})}. \bibinfo{publisher}{Association for
  Computing Machinery}, \bibinfo{address}{New York, NY, USA},
  \bibinfo{pages}{386–399}.
\newblock
\showISBNx{9781450349444}
\urldef\tempurl%
\url{https://doi.org/10.1145/3052973.3053032}
\showDOI{\tempurl}


\bibitem[\protect\citeauthoryear{Egelman, Jain, Portnoff, Liao, Consolvo, and
  Wagner}{Egelman et~al\mbox{.}}{2014}]%
        {ready-to-lock}
\bibfield{author}{\bibinfo{person}{Serge Egelman}, \bibinfo{person}{Sakshi
  Jain}, \bibinfo{person}{Rebecca~S. Portnoff}, \bibinfo{person}{Kerwell Liao},
  \bibinfo{person}{Sunny Consolvo}, {and} \bibinfo{person}{David Wagner}.}
  \bibinfo{year}{2014}\natexlab{}.
\newblock \showarticletitle{Are You Ready to Lock?}. In
  \bibinfo{booktitle}{\emph{Proceedings of the 2014 ACM SIGSAC Conference on
  Computer and Communications Security}} (Scottsdale, Arizona, USA)
  \emph{(\bibinfo{series}{CCS '14})}. \bibinfo{publisher}{ACM},
  \bibinfo{address}{New York, NY, USA}, \bibinfo{pages}{750--761}.
\newblock
\showISBNx{978-1-4503-2957-6}
\urldef\tempurl%
\url{https://doi.org/10.1145/2660267.2660273}
\showDOI{\tempurl}


\bibitem[\protect\citeauthoryear{Ehatisham-Ul-Haq, {Awais Azam}, Naeem, Amin,
  and Loo}{Ehatisham-Ul-Haq et~al\mbox{.}}{2018}]%
        {only-sensors}
\bibfield{author}{\bibinfo{person}{Muhammad Ehatisham-Ul-Haq},
  \bibinfo{person}{Muhammad {Awais Azam}}, \bibinfo{person}{Usman Naeem},
  \bibinfo{person}{Yasar Amin}, {and} \bibinfo{person}{Jonathan Loo}.}
  \bibinfo{year}{2018}\natexlab{}.
\newblock \showarticletitle{Continuous authentication of smartphone users based
  on activity pattern recognition using passive mobile sensing}.
\newblock \bibinfo{journal}{\emph{Journal of Network and Computer
  Applications}}  \bibinfo{volume}{109} (\bibinfo{year}{2018}),
  \bibinfo{pages}{24--35}.
\newblock
\showISSN{1084-8045}
\urldef\tempurl%
\url{https://doi.org/10.1016/j.jnca.2018.02.020}
\showDOI{\tempurl}


\bibitem[\protect\citeauthoryear{Ellavarason, Guest, Deravi, Sanchez-Riello,
  and Corsetti}{Ellavarason et~al\mbox{.}}{2020}]%
        {overview-3}
\bibfield{author}{\bibinfo{person}{Elakkiya Ellavarason},
  \bibinfo{person}{Richard Guest}, \bibinfo{person}{Farzin Deravi},
  \bibinfo{person}{Raul Sanchez-Riello}, {and} \bibinfo{person}{Barbara
  Corsetti}.} \bibinfo{year}{2020}\natexlab{}.
\newblock \showarticletitle{Touch-Dynamics Based Behavioural Biometrics on
  Mobile Devices – A Review from a Usability and Performance Perspective}.
\newblock \bibinfo{journal}{\emph{ACM Comput. Surv.}} \bibinfo{volume}{53},
  \bibinfo{number}{6}, Article \bibinfo{articleno}{120} (\bibinfo{date}{Dec.}
  \bibinfo{year}{2020}), \bibinfo{numpages}{36}~pages.
\newblock
\showISSN{0360-0300}
\urldef\tempurl%
\url{https://doi.org/10.1145/3394713}
\showDOI{\tempurl}


\bibitem[\protect\citeauthoryear{Fathy, Patel, and Chellappa}{Fathy
  et~al\mbox{.}}{2015}]%
        {cont-face}
\bibfield{author}{\bibinfo{person}{Mohammed~E Fathy}, \bibinfo{person}{Vishal~M
  Patel}, {and} \bibinfo{person}{Rama Chellappa}.}
  \bibinfo{year}{2015}\natexlab{}.
\newblock \showarticletitle{Face-based active authentication on mobile
  devices}. In \bibinfo{booktitle}{\emph{2015 IEEE international conference on
  acoustics, speech and signal processing (ICASSP)}}. IEEE,
  \bibinfo{pages}{1687--1691}.
\newblock


\bibitem[\protect\citeauthoryear{Feng, Liu, Kwon, Shi, Carbunar, Jiang, and
  Nguyen}{Feng et~al\mbox{.}}{2012}]%
        {continuous-mobile-touchscreen-gestures}
\bibfield{author}{\bibinfo{person}{Tao Feng}, \bibinfo{person}{Ziyi Liu},
  \bibinfo{person}{Kyeong-An Kwon}, \bibinfo{person}{Weidong Shi},
  \bibinfo{person}{Bogdan Carbunar}, \bibinfo{person}{Yifei Jiang}, {and}
  \bibinfo{person}{Nhung Nguyen}.} \bibinfo{year}{2012}\natexlab{}.
\newblock \showarticletitle{Continuous mobile authentication using touchscreen
  gestures}. In \bibinfo{booktitle}{\emph{2012 IEEE Conference on Technologies
  for Homeland Security (HST)}}. \bibinfo{pages}{451--456}.
\newblock
\urldef\tempurl%
\url{https://doi.org/10.1109/THS.2012.6459891}
\showDOI{\tempurl}


\bibitem[\protect\citeauthoryear{Feng, Yang, Yan, Tapia, and Shi}{Feng
  et~al\mbox{.}}{2014}]%
        {tips}
\bibfield{author}{\bibinfo{person}{Tao Feng}, \bibinfo{person}{Jun Yang},
  \bibinfo{person}{Zhixian Yan}, \bibinfo{person}{Emmanuel~Munguia Tapia},
  {and} \bibinfo{person}{Weidong Shi}.} \bibinfo{year}{2014}\natexlab{}.
\newblock \showarticletitle{TIPS: Context-Aware Implicit User Identification
  Using Touch Screen in Uncontrolled Environments}. In
  \bibinfo{booktitle}{\emph{Proceedings of the 15th Workshop on Mobile
  Computing Systems and Applications}} (Santa Barbara, California)
  \emph{(\bibinfo{series}{HotMobile '14})}. \bibinfo{publisher}{Association for
  Computing Machinery}, \bibinfo{address}{New York, NY, USA}, Article
  \bibinfo{articleno}{9}, \bibinfo{numpages}{6}~pages.
\newblock
\showISBNx{9781450327428}
\urldef\tempurl%
\url{https://doi.org/10.1145/2565585.2565592}
\showDOI{\tempurl}


\bibitem[\protect\citeauthoryear{Fierrez, Pozo, Martinez-Diaz, Galbally, and
  Morales}{Fierrez et~al\mbox{.}}{2018}]%
        {overview-2}
\bibfield{author}{\bibinfo{person}{Julian Fierrez}, \bibinfo{person}{Ada Pozo},
  \bibinfo{person}{Marcos Martinez-Diaz}, \bibinfo{person}{Javier Galbally},
  {and} \bibinfo{person}{Aythami Morales}.} \bibinfo{year}{2018}\natexlab{}.
\newblock \showarticletitle{Benchmarking Touchscreen Biometrics for Mobile
  Authentication}.
\newblock \bibinfo{journal}{\emph{IEEE Transactions on Information Forensics
  and Security}} \bibinfo{volume}{13}, \bibinfo{number}{11}
  (\bibinfo{year}{2018}), \bibinfo{pages}{2720--2733}.
\newblock
\urldef\tempurl%
\url{https://doi.org/10.1109/TIFS.2018.2833042}
\showDOI{\tempurl}


\bibitem[\protect\citeauthoryear{Filippov, Iuzbashev, and Kurnev}{Filippov
  et~al\mbox{.}}{2018}]%
        {touch-pattern-isolation}
\bibfield{author}{\bibinfo{person}{Alexander~I. Filippov},
  \bibinfo{person}{Artem~V. Iuzbashev}, {and} \bibinfo{person}{Alexey~S.
  Kurnev}.} \bibinfo{year}{2018}\natexlab{}.
\newblock \showarticletitle{User authentication via touch pattern recognition
  based on isolation forest}. In \bibinfo{booktitle}{\emph{2018 IEEE Conference
  of Russian Young Researchers in Electrical and Electronic Engineering
  (EIConRus)}}. \bibinfo{pages}{1485--1489}.
\newblock
\urldef\tempurl%
\url{https://doi.org/10.1109/EIConRus.2018.8317378}
\showDOI{\tempurl}


\bibitem[\protect\citeauthoryear{{Frank}, {Biedert}, {Ma}, {Martinovic}, and
  {Song}}{{Frank} et~al\mbox{.}}{2013}]%
        {touchalytics}
\bibfield{author}{\bibinfo{person}{M. {Frank}}, \bibinfo{person}{R. {Biedert}},
  \bibinfo{person}{E. {Ma}}, \bibinfo{person}{I. {Martinovic}}, {and}
  \bibinfo{person}{D. {Song}}.} \bibinfo{year}{2013}\natexlab{}.
\newblock \showarticletitle{Touchalytics: On the Applicability of Touchscreen
  Input as a Behavioral Biometric for Continuous Authentication}.
\newblock \bibinfo{journal}{\emph{IEEE Transactions on Information Forensics
  and Security}} \bibinfo{volume}{8}, \bibinfo{number}{1}
  (\bibinfo{year}{2013}), \bibinfo{pages}{136--148}.
\newblock


\bibitem[\protect\citeauthoryear{Galbally, Fierrez, Alonso-Fernandez, and
  Martinez-Diaz}{Galbally et~al\mbox{.}}{2011}]%
        {fingerprint-attack-1}
\bibfield{author}{\bibinfo{person}{J. Galbally}, \bibinfo{person}{J. Fierrez},
  \bibinfo{person}{F. Alonso-Fernandez}, {and} \bibinfo{person}{M.
  Martinez-Diaz}.} \bibinfo{year}{2011}\natexlab{}.
\newblock \showarticletitle{Evaluation of direct attacks to fingerprint
  verification systems}.
\newblock \bibinfo{journal}{\emph{Telecommunication Systems}}
  \bibinfo{volume}{47}, \bibinfo{number}{3} (\bibinfo{date}{01 Aug}
  \bibinfo{year}{2011}), \bibinfo{pages}{243--254}.
\newblock
\showISSN{1572-9451}
\urldef\tempurl%
\url{https://doi.org/10.1007/s11235-010-9316-0}
\showDOI{\tempurl}


\bibitem[\protect\citeauthoryear{Gascon, Uellenbeck, Wolf, and Rieck}{Gascon
  et~al\mbox{.}}{2014}]%
        {cont-typing}
\bibfield{author}{\bibinfo{person}{Hugo Gascon}, \bibinfo{person}{Sebastian
  Uellenbeck}, \bibinfo{person}{Christopher Wolf}, {and}
  \bibinfo{person}{Konrad Rieck}.} \bibinfo{year}{2014}\natexlab{}.
\newblock \showarticletitle{Continuous authentication on mobile devices by
  analysis of typing motion behavior}.
\newblock \bibinfo{journal}{\emph{Sicherheit 2014--Sicherheit, Schutz und
  Zuverl{\"a}ssigkeit}} (\bibinfo{year}{2014}).
\newblock


\bibitem[\protect\citeauthoryear{Georgiev, Eberz, Turner, Lovisotto, and
  Martinovic}{Georgiev et~al\mbox{.}}{2022}]%
        {common-evaluation}
\bibfield{author}{\bibinfo{person}{Martin Georgiev}, \bibinfo{person}{Simon
  Eberz}, \bibinfo{person}{Henry Turner}, \bibinfo{person}{Giulio Lovisotto},
  {and} \bibinfo{person}{Ivan Martinovic}.} \bibinfo{year}{2022}\natexlab{}.
\newblock \showarticletitle{Common Evaluation Pitfalls in Touch-Based
  Authentication Systems}. In \bibinfo{booktitle}{\emph{Proceedings of the 2022
  ACM on Asia Conference on Computer and Communications Security}} (Nagasaki,
  Japan) \emph{(\bibinfo{series}{ASIA CCS '22})}.
  \bibinfo{publisher}{Association for Computing Machinery},
  \bibinfo{address}{New York, NY, USA}, \bibinfo{pages}{1049–1063}.
\newblock
\showISBNx{9781450391405}
\urldef\tempurl%
\url{https://doi.org/10.1145/3488932.3517388}
\showDOI{\tempurl}


\bibitem[\protect\citeauthoryear{Incel, Günay, Akan, Barlas, Basar, Alptekin,
  and Isbilen}{Incel et~al\mbox{.}}{2021}]%
        {dakota}
\bibfield{author}{\bibinfo{person}{Özlem~Durmaz Incel},
  \bibinfo{person}{Seçil Günay}, \bibinfo{person}{Yasemin Akan},
  \bibinfo{person}{Yunus Barlas}, \bibinfo{person}{Okan~Engin Basar},
  \bibinfo{person}{Gülfem~Isiklar Alptekin}, {and} \bibinfo{person}{Mustafa
  Isbilen}.} \bibinfo{year}{2021}\natexlab{}.
\newblock \showarticletitle{DAKOTA: Sensor and Touch Screen-Based Continuous
  Authentication on a Mobile Banking Application}.
\newblock \bibinfo{journal}{\emph{IEEE Access}}  \bibinfo{volume}{9}
  (\bibinfo{year}{2021}), \bibinfo{pages}{38943--38960}.
\newblock
\urldef\tempurl%
\url{https://doi.org/10.1109/ACCESS.2021.3063424}
\showDOI{\tempurl}


\bibitem[\protect\citeauthoryear{Jorgensen and Yu}{Jorgensen and Yu}{2011}]%
        {mouse-movement}
\bibfield{author}{\bibinfo{person}{Zach Jorgensen} {and} \bibinfo{person}{Ting
  Yu}.} \bibinfo{year}{2011}\natexlab{}.
\newblock \showarticletitle{On Mouse Dynamics as a Behavioral Biometric for
  Authentication}. In \bibinfo{booktitle}{\emph{Proceedings of the 6th ACM
  Symposium on Information, Computer and Communications Security}} (Hong Kong,
  China) \emph{(\bibinfo{series}{ASIACCS '11})}.
  \bibinfo{publisher}{Association for Computing Machinery},
  \bibinfo{address}{New York, NY, USA}, \bibinfo{pages}{476–482}.
\newblock
\showISBNx{9781450305648}
\urldef\tempurl%
\url{https://doi.org/10.1145/1966913.1966983}
\showDOI{\tempurl}


\bibitem[\protect\citeauthoryear{Kim and Kang}{Kim and Kang}{2020}]%
        {keystroke-mobile}
\bibfield{author}{\bibinfo{person}{Junhong Kim} {and} \bibinfo{person}{Pilsung
  Kang}.} \bibinfo{year}{2020}\natexlab{}.
\newblock \showarticletitle{Freely typed keystroke dynamics-based user
  authentication for mobile devices based on heterogeneous features}.
\newblock \bibinfo{journal}{\emph{Pattern Recognition}}  \bibinfo{volume}{108}
  (\bibinfo{year}{2020}), \bibinfo{pages}{107556}.
\newblock
\showISSN{0031-3203}
\urldef\tempurl%
\url{https://doi.org/10.1016/j.patcog.2020.107556}
\showDOI{\tempurl}


\bibitem[\protect\citeauthoryear{Kumar, Kundu, and Phoha}{Kumar
  et~al\mbox{.}}{2018}]%
        {one-class}
\bibfield{author}{\bibinfo{person}{Rajesh Kumar},
  \bibinfo{person}{Partha~Pratim Kundu}, {and} \bibinfo{person}{Vir~V. Phoha}.}
  \bibinfo{year}{2018}\natexlab{}.
\newblock \showarticletitle{Continuous authentication using one-class
  classifiers and their fusion}. In \bibinfo{booktitle}{\emph{2018 IEEE 4th
  International Conference on Identity, Security, and Behavior Analysis
  (ISBA)}}. \bibinfo{pages}{1--8}.
\newblock
\urldef\tempurl%
\url{https://doi.org/10.1109/ISBA.2018.8311467}
\showDOI{\tempurl}


\bibitem[\protect\citeauthoryear{Kumar, Phoha, and Serwadda}{Kumar
  et~al\mbox{.}}{2016}]%
        {fusing-swiping}
\bibfield{author}{\bibinfo{person}{Rajesh Kumar}, \bibinfo{person}{Vir~V.
  Phoha}, {and} \bibinfo{person}{Abdul Serwadda}.}
  \bibinfo{year}{2016}\natexlab{}.
\newblock \showarticletitle{Continuous authentication of smartphone users by
  fusing typing, swiping, and phone movement patterns}. In
  \bibinfo{booktitle}{\emph{2016 IEEE 8th International Conference on
  Biometrics Theory, Applications and Systems (BTAS)}}. \bibinfo{pages}{1--8}.
\newblock
\urldef\tempurl%
\url{https://doi.org/10.1109/BTAS.2016.7791164}
\showDOI{\tempurl}


\bibitem[\protect\citeauthoryear{Li, Zhao, and Xue}{Li et~al\mbox{.}}{2013}]%
        {unobservable-re-authentication}
\bibfield{author}{\bibinfo{person}{Lingjun Li}, \bibinfo{person}{Xinxin Zhao},
  {and} \bibinfo{person}{Guoliang Xue}.} \bibinfo{year}{2013}\natexlab{}.
\newblock \showarticletitle{Unobservable Re-authentication for Smartphones}. In
  \bibinfo{booktitle}{\emph{20th Annual Network and Distributed System Security
  Symposium, {NDSS} 2013, San Diego, California, USA, February 24-27, 2013}}.
  \bibinfo{publisher}{The Internet Society}.
\newblock
\urldef\tempurl%
\url{https://www.ndss-symposium.org/ndss2013/unobservable-re-authentication-smartphones}
\showURL{%
\tempurl}


\bibitem[\protect\citeauthoryear{Mahbub, Sarkar, Patel, and Chellappa}{Mahbub
  et~al\mbox{.}}{2016}]%
        {active-user-authentication}
\bibfield{author}{\bibinfo{person}{Upal Mahbub}, \bibinfo{person}{Sayantan
  Sarkar}, \bibinfo{person}{Vishal~M. Patel}, {and} \bibinfo{person}{Rama
  Chellappa}.} \bibinfo{year}{2016}\natexlab{}.
\newblock \showarticletitle{Active user authentication for smartphones: A
  challenge data set and benchmark results}. In \bibinfo{booktitle}{\emph{2016
  IEEE 8th International Conference on Biometrics Theory, Applications and
  Systems (BTAS)}}. \bibinfo{pages}{1--8}.
\newblock
\urldef\tempurl%
\url{https://doi.org/10.1109/BTAS.2016.7791155}
\showDOI{\tempurl}


\bibitem[\protect\citeauthoryear{Meng, Wang, Wong, Wen, and Xiang}{Meng
  et~al\mbox{.}}{2018}]%
        {touch-wb}
\bibfield{author}{\bibinfo{person}{Weizhi Meng}, \bibinfo{person}{Yu Wang},
  \bibinfo{person}{Duncan~S. Wong}, \bibinfo{person}{Sheng Wen}, {and}
  \bibinfo{person}{Yang Xiang}.} \bibinfo{year}{2018}\natexlab{}.
\newblock \showarticletitle{TouchWB: Touch behavioral user authentication based
  on web browsing on smartphones}.
\newblock \bibinfo{journal}{\emph{Journal of Network and Computer
  Applications}}  \bibinfo{volume}{117} (\bibinfo{year}{2018}),
  \bibinfo{pages}{1--9}.
\newblock
\showISSN{1084-8045}
\urldef\tempurl%
\url{https://doi.org/10.1016/j.jnca.2018.05.010}
\showDOI{\tempurl}


\bibitem[\protect\citeauthoryear{Meng, Wong, and Kwok}{Meng
  et~al\mbox{.}}{2014}]%
        {design-of-touch}
\bibfield{author}{\bibinfo{person}{Yuxin Meng}, \bibinfo{person}{Duncan~S.
  Wong}, {and} \bibinfo{person}{Lam-For Kwok}.}
  \bibinfo{year}{2014}\natexlab{}.
\newblock \showarticletitle{Design of Touch Dynamics Based User Authentication
  with an Adaptive Mechanism on Mobile Phones}. In
  \bibinfo{booktitle}{\emph{Proceedings of the 29th Annual ACM Symposium on
  Applied Computing}} (Gyeongju, Republic of Korea) \emph{(\bibinfo{series}{SAC
  '14})}. \bibinfo{publisher}{Association for Computing Machinery},
  \bibinfo{address}{New York, NY, USA}, \bibinfo{pages}{1680–1687}.
\newblock
\showISBNx{9781450324694}
\urldef\tempurl%
\url{https://doi.org/10.1145/2554850.2554931}
\showDOI{\tempurl}


\bibitem[\protect\citeauthoryear{Moher, Liberati, Tetzlaff, Altman, and
  Group}{Moher et~al\mbox{.}}{2009}]%
        {moher2009preferred}
\bibfield{author}{\bibinfo{person}{David Moher}, \bibinfo{person}{Alessandro
  Liberati}, \bibinfo{person}{Jennifer Tetzlaff}, \bibinfo{person}{Douglas~G
  Altman}, {and} \bibinfo{person}{Prisma Group}.}
  \bibinfo{year}{2009}\natexlab{}.
\newblock \showarticletitle{Preferred reporting items for systematic reviews
  and meta-analyses: the PRISMA statement}.
\newblock \bibinfo{journal}{\emph{PLoS medicine}} \bibinfo{volume}{6},
  \bibinfo{number}{7} (\bibinfo{year}{2009}), \bibinfo{pages}{e1000097}.
\newblock


\bibitem[\protect\citeauthoryear{Mondal and Bours}{Mondal and Bours}{2015a}]%
        {trust-model}
\bibfield{author}{\bibinfo{person}{Soumik Mondal} {and}
  \bibinfo{person}{Patrick Bours}.} \bibinfo{year}{2015}\natexlab{a}.
\newblock \showarticletitle{A computational approach to the continuous
  authentication biometric system}.
\newblock \bibinfo{journal}{\emph{Information Sciences}}  \bibinfo{volume}{304}
  (\bibinfo{year}{2015}), \bibinfo{pages}{28--53}.
\newblock
\showISSN{0020-0255}
\urldef\tempurl%
\url{https://doi.org/10.1016/j.ins.2014.12.045}
\showDOI{\tempurl}


\bibitem[\protect\citeauthoryear{Mondal and Bours}{Mondal and Bours}{2015b}]%
        {swipe-gesture}
\bibfield{author}{\bibinfo{person}{Soumik Mondal} {and}
  \bibinfo{person}{Patrick Bours}.} \bibinfo{year}{2015}\natexlab{b}.
\newblock \showarticletitle{Swipe gesture based Continuous Authentication for
  mobile devices}. In \bibinfo{booktitle}{\emph{2015 International Conference
  on Biometrics (ICB)}}. \bibinfo{pages}{458--465}.
\newblock
\urldef\tempurl%
\url{https://doi.org/10.1109/ICB.2015.7139110}
\showDOI{\tempurl}


\bibitem[\protect\citeauthoryear{Murmuria, Stavrou, Barbar{\'a}, and
  Fleck}{Murmuria et~al\mbox{.}}{2015}]%
        {power-consumption}
\bibfield{author}{\bibinfo{person}{Rahul Murmuria}, \bibinfo{person}{Angelos
  Stavrou}, \bibinfo{person}{Daniel Barbar{\'a}}, {and} \bibinfo{person}{Dan
  Fleck}.} \bibinfo{year}{2015}\natexlab{}.
\newblock \showarticletitle{Continuous Authentication on Mobile Devices Using
  Power Consumption, Touch Gestures and Physical Movement of Users}. In
  \bibinfo{booktitle}{\emph{Research in Attacks, Intrusions, and Defenses}},
  \bibfield{editor}{\bibinfo{person}{Herbert Bos}, \bibinfo{person}{Fabian
  Monrose}, {and} \bibinfo{person}{Gregory Blanc}} (Eds.).
  \bibinfo{publisher}{Springer International Publishing},
  \bibinfo{address}{Cham}, \bibinfo{pages}{405--424}.
\newblock
\showISBNx{978-3-319-26362-5}


\bibitem[\protect\citeauthoryear{Papamichail, Chatzidimitriou, Karanikiotis,
  Oikonomou, Symeonidis, and Saripalle}{Papamichail et~al\mbox{.}}{2019}]%
        {brainrun}
\bibfield{author}{\bibinfo{person}{Michail~D. Papamichail},
  \bibinfo{person}{Kyriakos~C. Chatzidimitriou}, \bibinfo{person}{Thomas
  Karanikiotis}, \bibinfo{person}{Napoleon-Christos~I. Oikonomou},
  \bibinfo{person}{Andreas~L. Symeonidis}, {and} \bibinfo{person}{Sashi~K.
  Saripalle}.} \bibinfo{year}{2019}\natexlab{}.
\newblock \showarticletitle{BrainRun: A Behavioral Biometrics Dataset towards
  Continuous Implicit Authentication}.
\newblock \bibinfo{journal}{\emph{Data}} \bibinfo{volume}{4},
  \bibinfo{number}{2} (\bibinfo{year}{2019}).
\newblock
\showISSN{2306-5729}
\urldef\tempurl%
\url{https://doi.org/10.3390/data4020060}
\showDOI{\tempurl}


\bibitem[\protect\citeauthoryear{Patel, Chellappa, Chandra, and Barbello}{Patel
  et~al\mbox{.}}{2016}]%
        {overview-1}
\bibfield{author}{\bibinfo{person}{Vishal~M. Patel}, \bibinfo{person}{Rama
  Chellappa}, \bibinfo{person}{Deepak Chandra}, {and} \bibinfo{person}{Brandon
  Barbello}.} \bibinfo{year}{2016}\natexlab{}.
\newblock \showarticletitle{Continuous User Authentication on Mobile Devices:
  Recent progress and remaining challenges}.
\newblock \bibinfo{journal}{\emph{IEEE Signal Processing Magazine}}
  \bibinfo{volume}{33}, \bibinfo{number}{4} (\bibinfo{year}{2016}),
  \bibinfo{pages}{49--61}.
\newblock
\urldef\tempurl%
\url{https://doi.org/10.1109/MSP.2016.2555335}
\showDOI{\tempurl}


\bibitem[\protect\citeauthoryear{Ramachandra and Busch}{Ramachandra and
  Busch}{2017}]%
        {face-recognition-attack}
\bibfield{author}{\bibinfo{person}{Raghavendra Ramachandra} {and}
  \bibinfo{person}{Christoph Busch}.} \bibinfo{year}{2017}\natexlab{}.
\newblock \showarticletitle{Presentation Attack Detection Methods for Face
  Recognition Systems: A Comprehensive Survey}.
\newblock \bibinfo{journal}{\emph{ACM Comput. Surv.}} \bibinfo{volume}{50},
  \bibinfo{number}{1}, Article \bibinfo{articleno}{8} (\bibinfo{date}{March}
  \bibinfo{year}{2017}), \bibinfo{numpages}{37}~pages.
\newblock
\showISSN{0360-0300}
\urldef\tempurl%
\url{https://doi.org/10.1145/3038924}
\showDOI{\tempurl}


\bibitem[\protect\citeauthoryear{Rasnayaka and Sim}{Rasnayaka and Sim}{2018}]%
        {continuous-support}
\bibfield{author}{\bibinfo{person}{Sanka Rasnayaka} {and}
  \bibinfo{person}{Terence Sim}.} \bibinfo{year}{2018}\natexlab{}.
\newblock \showarticletitle{Who wants Continuous Authentication on Mobile
  Devices?}. In \bibinfo{booktitle}{\emph{2018 IEEE 9th International
  Conference on Biometrics Theory, Applications and Systems (BTAS)}}.
  \bibinfo{pages}{1--9}.
\newblock
\urldef\tempurl%
\url{https://doi.org/10.1109/BTAS.2018.8698599}
\showDOI{\tempurl}


\bibitem[\protect\citeauthoryear{Rocha, Carneiro, and Novais}{Rocha
  et~al\mbox{.}}{2021}]%
        {explainability}
\bibfield{author}{\bibinfo{person}{Rodrigo Rocha}, \bibinfo{person}{Davide
  Carneiro}, {and} \bibinfo{person}{Paulo Novais}.}
  \bibinfo{year}{2021}\natexlab{}.
\newblock \showarticletitle{Continuous authentication with a focus on
  explainability}.
\newblock \bibinfo{journal}{\emph{Neurocomputing}}  \bibinfo{volume}{423}
  (\bibinfo{year}{2021}), \bibinfo{pages}{697--702}.
\newblock
\showISSN{0925-2312}
\urldef\tempurl%
\url{https://doi.org/10.1016/j.neucom.2020.02.122}
\showDOI{\tempurl}


\bibitem[\protect\citeauthoryear{Roy, Halevi, and Memon}{Roy
  et~al\mbox{.}}{2014}]%
        {hmm-model}
\bibfield{author}{\bibinfo{person}{Aditi Roy}, \bibinfo{person}{Tzipora
  Halevi}, {and} \bibinfo{person}{Nasir Memon}.}
  \bibinfo{year}{2014}\natexlab{}.
\newblock \showarticletitle{An HMM-based behavior modeling approach for
  continuous mobile authentication}. In \bibinfo{booktitle}{\emph{2014 IEEE
  International Conference on Acoustics, Speech and Signal Processing
  (ICASSP)}}. \bibinfo{pages}{3789--3793}.
\newblock
\urldef\tempurl%
\url{https://doi.org/10.1109/ICASSP.2014.6854310}
\showDOI{\tempurl}


\bibitem[\protect\citeauthoryear{Samet, Ishraque, Ghadamyari, Kakadiya, Mistry,
  and Nakkabi}{Samet et~al\mbox{.}}{2019}]%
        {touch-metric}
\bibfield{author}{\bibinfo{person}{Saeed Samet}, \bibinfo{person}{Mohd~Tazim
  Ishraque}, \bibinfo{person}{Mehdi Ghadamyari}, \bibinfo{person}{Krishna
  Kakadiya}, \bibinfo{person}{Yash Mistry}, {and} \bibinfo{person}{Youssef
  Nakkabi}.} \bibinfo{year}{2019}\natexlab{}.
\newblock \showarticletitle{TouchMetric: a machine learning based continuous
  authentication feature testing mobile application}.
\newblock \bibinfo{journal}{\emph{International Journal of Information
  Technology}} \bibinfo{volume}{11}, \bibinfo{number}{4} (\bibinfo{date}{01
  Dec} \bibinfo{year}{2019}), \bibinfo{pages}{625--631}.
\newblock
\showISSN{2511-2112}
\urldef\tempurl%
\url{https://doi.org/10.1007/s41870-019-00306-w}
\showDOI{\tempurl}


\bibitem[\protect\citeauthoryear{Saravanan, Clarke, Chau, and Zha}{Saravanan
  et~al\mbox{.}}{2014}]%
        {latent-gestures}
\bibfield{author}{\bibinfo{person}{Premkumar Saravanan},
  \bibinfo{person}{Samuel Clarke}, \bibinfo{person}{Duen Horng~(Polo) Chau},
  {and} \bibinfo{person}{Hongyuan Zha}.} \bibinfo{year}{2014}\natexlab{}.
\newblock \showarticletitle{LatentGesture: Active User Authentication through
  Background Touch Analysis}. In \bibinfo{booktitle}{\emph{Proceedings of the
  Second International Symposium of Chinese CHI}} (Toronto, Ontario, Canada)
  \emph{(\bibinfo{series}{Chinese CHI '14})}. \bibinfo{publisher}{Association
  for Computing Machinery}, \bibinfo{address}{New York, NY, USA},
  \bibinfo{pages}{110–113}.
\newblock
\showISBNx{9781450328760}
\urldef\tempurl%
\url{https://doi.org/10.1145/2592235.2592252}
\showDOI{\tempurl}


\bibitem[\protect\citeauthoryear{{Serwadda}, {Phoha}, and {Wang}}{{Serwadda}
  et~al\mbox{.}}{2013}]%
        {which-verifiers-work}
\bibfield{author}{\bibinfo{person}{A. {Serwadda}}, \bibinfo{person}{V.~V.
  {Phoha}}, {and} \bibinfo{person}{Z. {Wang}}.}
  \bibinfo{year}{2013}\natexlab{}.
\newblock \showarticletitle{Which verifiers work?: A benchmark evaluation of
  touch-based authentication algorithms}. In \bibinfo{booktitle}{\emph{2013
  IEEE Sixth International Conference on Biometrics: Theory, Applications and
  Systems (BTAS)}}. \bibinfo{pages}{1--8}.
\newblock
\urldef\tempurl%
\url{https://doi.org/10.1109/BTAS.2013.6712758}
\showDOI{\tempurl}


\bibitem[\protect\citeauthoryear{Shen, Zhang, Guan, and Maxion}{Shen
  et~al\mbox{.}}{2016}]%
        {performance-analysis}
\bibfield{author}{\bibinfo{person}{Chao Shen}, \bibinfo{person}{Yong Zhang},
  \bibinfo{person}{Xiaohong Guan}, {and} \bibinfo{person}{Roy~A. Maxion}.}
  \bibinfo{year}{2016}\natexlab{}.
\newblock \showarticletitle{Performance Analysis of Touch-Interaction Behavior
  for Active Smartphone Authentication}.
\newblock \bibinfo{journal}{\emph{IEEE Transactions on Information Forensics
  and Security}} \bibinfo{volume}{11}, \bibinfo{number}{3}
  (\bibinfo{year}{2016}), \bibinfo{pages}{498--513}.
\newblock
\urldef\tempurl%
\url{https://doi.org/10.1109/TIFS.2015.2503258}
\showDOI{\tempurl}


\bibitem[\protect\citeauthoryear{Sitová, Šeděnka, Yang, Peng, Zhou, Gasti,
  and Balagani}{Sitová et~al\mbox{.}}{2016}]%
        {hmog-sensors}
\bibfield{author}{\bibinfo{person}{Zdeňka Sitová}, \bibinfo{person}{Jaroslav
  Šeděnka}, \bibinfo{person}{Qing Yang}, \bibinfo{person}{Ge Peng},
  \bibinfo{person}{Gang Zhou}, \bibinfo{person}{Paolo Gasti}, {and}
  \bibinfo{person}{Kiran~S. Balagani}.} \bibinfo{year}{2016}\natexlab{}.
\newblock \showarticletitle{HMOG: New Behavioral Biometric Features for
  Continuous Authentication of Smartphone Users}.
\newblock \bibinfo{journal}{\emph{IEEE Transactions on Information Forensics
  and Security}} \bibinfo{volume}{11}, \bibinfo{number}{5}
  (\bibinfo{year}{2016}), \bibinfo{pages}{877--892}.
\newblock
\urldef\tempurl%
\url{https://doi.org/10.1109/TIFS.2015.2506542}
\showDOI{\tempurl}


\bibitem[\protect\citeauthoryear{Song, Cai, and Zhang}{Song
  et~al\mbox{.}}{2017}]%
        {multitouch-only}
\bibfield{author}{\bibinfo{person}{Yunpeng Song}, \bibinfo{person}{Zhongmin
  Cai}, {and} \bibinfo{person}{Zhi-Li Zhang}.} \bibinfo{year}{2017}\natexlab{}.
\newblock \showarticletitle{Multi-touch Authentication Using Hand Geometry and
  Behavioral Information}. In \bibinfo{booktitle}{\emph{2017 IEEE Symposium on
  Security and Privacy (SP)}}. \bibinfo{pages}{357--372}.
\newblock
\urldef\tempurl%
\url{https://doi.org/10.1109/SP.2017.54}
\showDOI{\tempurl}


\bibitem[\protect\citeauthoryear{Sugrim, Liu, McLean, and Lindqvist}{Sugrim
  et~al\mbox{.}}{[n.d.]}]%
        {robust-performance}
\bibfield{author}{\bibinfo{person}{Shridatt Sugrim}, \bibinfo{person}{Can Liu},
  \bibinfo{person}{Meghan McLean}, {and} \bibinfo{person}{Janne Lindqvist}.}
  \bibinfo{year}{[n.d.]}\natexlab{}.
\newblock \showarticletitle{Robust Performance Metrics for Authentication
  Systems}.
\newblock \bibinfo{journal}{\emph{Network and Distributed Systems Security
  (NDSS) Symposium 2019}} (\bibinfo{year}{[n.\,d.]}).
\newblock
\urldef\tempurl%
\url{https://doi.org/10.14722/ndss.2019.23351}
\showDOI{\tempurl}


\bibitem[\protect\citeauthoryear{Syed, Helmick, Banerjee, and Cukic}{Syed
  et~al\mbox{.}}{2019}]%
        {posture-size-config}
\bibfield{author}{\bibinfo{person}{Zahid Syed}, \bibinfo{person}{Jordan
  Helmick}, \bibinfo{person}{Sean Banerjee}, {and} \bibinfo{person}{Bojan
  Cukic}.} \bibinfo{year}{2019}\natexlab{}.
\newblock \showarticletitle{Touch gesture-based authentication on mobile
  devices: The effects of user posture, device size, configuration, and
  inter-session variability}.
\newblock \bibinfo{journal}{\emph{Journal of Systems and Software}}
  \bibinfo{volume}{149} (\bibinfo{year}{2019}), \bibinfo{pages}{158--173}.
\newblock
\showISSN{0164-1212}
\urldef\tempurl%
\url{https://doi.org/10.1016/j.jss.2018.11.017}
\showDOI{\tempurl}


\bibitem[\protect\citeauthoryear{Teh, Zhang, Teoh, and Chen}{Teh
  et~al\mbox{.}}{2016}]%
        {overview-4}
\bibfield{author}{\bibinfo{person}{Pin~Shen Teh}, \bibinfo{person}{Ning Zhang},
  \bibinfo{person}{Andrew Beng~Jin Teoh}, {and} \bibinfo{person}{Ke Chen}.}
  \bibinfo{year}{2016}\natexlab{}.
\newblock \showarticletitle{A survey on touch dynamics authentication in mobile
  devices}.
\newblock \bibinfo{journal}{\emph{Computers \& Security}}  \bibinfo{volume}{59}
  (\bibinfo{year}{2016}), \bibinfo{pages}{210--235}.
\newblock
\showISSN{0167-4048}
\urldef\tempurl%
\url{https://doi.org/10.1016/j.cose.2016.03.003}
\showDOI{\tempurl}


\bibitem[\protect\citeauthoryear{Volaka, Alptekin, Basar, Isbilen, and
  Incel}{Volaka et~al\mbox{.}}{2019}]%
        {cont-auth-deep-learning}
\bibfield{author}{\bibinfo{person}{Hasan~Can Volaka}, \bibinfo{person}{Gulfem
  Alptekin}, \bibinfo{person}{Okan~Engin Basar}, \bibinfo{person}{Mustafa
  Isbilen}, {and} \bibinfo{person}{Ozlem~Durmaz Incel}.}
  \bibinfo{year}{2019}\natexlab{}.
\newblock \showarticletitle{Towards Continuous Authentication on Mobile Phones
  using Deep Learning Models}.
\newblock \bibinfo{journal}{\emph{Procedia Computer Science}}
  \bibinfo{volume}{155} (\bibinfo{year}{2019}), \bibinfo{pages}{177--184}.
\newblock
\showISSN{1877-0509}
\urldef\tempurl%
\url{https://doi.org/10.1016/j.procs.2019.08.027}
\showDOI{\tempurl}
\newblock
\shownote{The 16th International Conference on Mobile Systems and Pervasive
  Computing (MobiSPC 2019),The 14th International Conference on Future Networks
  and Communications (FNC-2019),The 9th International Conference on Sustainable
  Energy Information Technology.}


\bibitem[\protect\citeauthoryear{Wang, Yu, Mengshoel, and Tague}{Wang
  et~al\mbox{.}}{2017}]%
        {towards-cont-passive}
\bibfield{author}{\bibinfo{person}{Xiao Wang}, \bibinfo{person}{Tong Yu},
  \bibinfo{person}{Ole Mengshoel}, {and} \bibinfo{person}{Patrick Tague}.}
  \bibinfo{year}{2017}\natexlab{}.
\newblock \showarticletitle{Towards Continuous and Passive Authentication
  across Mobile Devices: An Empirical Study}. In
  \bibinfo{booktitle}{\emph{Proceedings of the 10th ACM Conference on Security
  and Privacy in Wireless and Mobile Networks}} (Boston, Massachusetts)
  \emph{(\bibinfo{series}{WiSec '17})}. \bibinfo{publisher}{Association for
  Computing Machinery}, \bibinfo{address}{New York, NY, USA},
  \bibinfo{pages}{35–45}.
\newblock
\showISBNx{9781450350846}
\urldef\tempurl%
\url{https://doi.org/10.1145/3098243.3098244}
\showDOI{\tempurl}


\bibitem[\protect\citeauthoryear{Witten, Frank, Hall, Pal, and DATA}{Witten
  et~al\mbox{.}}{2005}]%
        {weka}
\bibfield{author}{\bibinfo{person}{Ian~H Witten}, \bibinfo{person}{Eibe Frank},
  \bibinfo{person}{Mark~A Hall}, \bibinfo{person}{Christopher~J Pal}, {and}
  \bibinfo{person}{MINING DATA}.} \bibinfo{year}{2005}\natexlab{}.
\newblock \showarticletitle{Practical machine learning tools and techniques}.
  In \bibinfo{booktitle}{\emph{DATA MINING}}, Vol.~\bibinfo{volume}{2}.
  \bibinfo{pages}{4}.
\newblock


\bibitem[\protect\citeauthoryear{Xu, Zhou, and Lyu}{Xu et~al\mbox{.}}{2014}]%
        {towards-continuous-passive}
\bibfield{author}{\bibinfo{person}{Hui Xu}, \bibinfo{person}{Yangfan Zhou},
  {and} \bibinfo{person}{Michael~R. Lyu}.} \bibinfo{year}{2014}\natexlab{}.
\newblock \showarticletitle{Towards Continuous and Passive Authentication via
  Touch Biometrics: An Experimental Study on Smartphones}. In
  \bibinfo{booktitle}{\emph{10th Symposium On Usable Privacy and Security
  ({SOUPS} 2014)}}. \bibinfo{publisher}{{USENIX} Association},
  \bibinfo{address}{Menlo Park, CA}, \bibinfo{pages}{187--198}.
\newblock
\showISBNx{978-1-931971-13-3}
\urldef\tempurl%
\url{https://www.usenix.org/conference/soups2014/proceedings/presentation/xu}
\showURL{%
\tempurl}


\bibitem[\protect\citeauthoryear{Xu, Yu, chen, Hua, Zhu, Chen, and Li}{Xu
  et~al\mbox{.}}{2020}]%
        {vibration}
\bibfield{author}{\bibinfo{person}{Xiangyu Xu}, \bibinfo{person}{Jiadi Yu},
  \bibinfo{person}{Yingying chen}, \bibinfo{person}{Qin Hua},
  \bibinfo{person}{Yanmin Zhu}, \bibinfo{person}{Yi-Chao Chen}, {and}
  \bibinfo{person}{Minglu Li}.} \bibinfo{year}{2020}\natexlab{}.
\newblock \showarticletitle{TouchPass: Towards Behavior-Irrelevant on-Touch
  User Authentication on Smartphones Leveraging Vibrations}. In
  \bibinfo{booktitle}{\emph{Proceedings of the 26th Annual International
  Conference on Mobile Computing and Networking}} (London, United Kingdom)
  \emph{(\bibinfo{series}{MobiCom '20})}. \bibinfo{publisher}{Association for
  Computing Machinery}, \bibinfo{address}{New York, NY, USA}, Article
  \bibinfo{articleno}{24}, \bibinfo{numpages}{13}~pages.
\newblock
\showISBNx{9781450370851}
\urldef\tempurl%
\url{https://doi.org/10.1145/3372224.3380901}
\showDOI{\tempurl}


\bibitem[\protect\citeauthoryear{Yang, Clark, Lindqvist, and Oulasvirta}{Yang
  et~al\mbox{.}}{2016}]%
        {free-form-gestures}
\bibfield{author}{\bibinfo{person}{Yulong Yang}, \bibinfo{person}{Gradeigh~D.
  Clark}, \bibinfo{person}{Janne Lindqvist}, {and} \bibinfo{person}{Antti
  Oulasvirta}.} \bibinfo{year}{2016}\natexlab{}.
\newblock \bibinfo{booktitle}{\emph{Free-Form Gesture Authentication in the
  Wild}}.
\newblock \bibinfo{publisher}{Association for Computing Machinery},
  \bibinfo{address}{New York, NY, USA}, \bibinfo{pages}{3722–3735}.
\newblock
\showISBNx{9781450333627}
\urldef\tempurl%
\url{https://doi.org/10.1145/2858036.2858270}
\showURL{%
\tempurl}


\bibitem[\protect\citeauthoryear{Yang, Guo, Wang, Li, Yu, and Zhou}{Yang
  et~al\mbox{.}}{2019}]%
        {behave-sense}
\bibfield{author}{\bibinfo{person}{Yafang Yang}, \bibinfo{person}{Bin Guo},
  \bibinfo{person}{Zhu Wang}, \bibinfo{person}{Mingyang Li},
  \bibinfo{person}{Zhiwen Yu}, {and} \bibinfo{person}{Xingshe Zhou}.}
  \bibinfo{year}{2019}\natexlab{}.
\newblock \showarticletitle{BehaveSense: Continuous authentication for
  security-sensitive mobile apps using behavioral biometrics}.
\newblock \bibinfo{journal}{\emph{Ad Hoc Networks}}  \bibinfo{volume}{84}
  (\bibinfo{year}{2019}), \bibinfo{pages}{9--18}.
\newblock
\showISSN{1570-8705}
\urldef\tempurl%
\url{https://doi.org/10.1016/j.adhoc.2018.09.015}
\showDOI{\tempurl}


\bibitem[\protect\citeauthoryear{Zhang, Patel, Fathy, and Chellappa}{Zhang
  et~al\mbox{.}}{2015}]%
        {dictionaries}
\bibfield{author}{\bibinfo{person}{Heng Zhang}, \bibinfo{person}{Vishal~M.
  Patel}, \bibinfo{person}{Mohammed Fathy}, {and} \bibinfo{person}{Rama
  Chellappa}.} \bibinfo{year}{2015}\natexlab{}.
\newblock \showarticletitle{Touch Gesture-Based Active User Authentication
  Using Dictionaries}. In \bibinfo{booktitle}{\emph{2015 IEEE Winter Conference
  on Applications of Computer Vision}}. \bibinfo{pages}{207--214}.
\newblock
\urldef\tempurl%
\url{https://doi.org/10.1109/WACV.2015.35}
\showDOI{\tempurl}


\bibitem[\protect\citeauthoryear{Zhao, Feng, and Shi}{Zhao
  et~al\mbox{.}}{2013}]%
        {graphic-feature}
\bibfield{author}{\bibinfo{person}{Xi Zhao}, \bibinfo{person}{Tao Feng}, {and}
  \bibinfo{person}{Weidong Shi}.} \bibinfo{year}{2013}\natexlab{}.
\newblock \showarticletitle{Continuous mobile authentication using a novel
  Graphic Touch Gesture Feature}. In \bibinfo{booktitle}{\emph{2013 IEEE Sixth
  International Conference on Biometrics: Theory, Applications and Systems
  (BTAS)}}. \bibinfo{pages}{1--6}.
\newblock
\urldef\tempurl%
\url{https://doi.org/10.1109/BTAS.2013.6712747}
\showDOI{\tempurl}


\bibitem[\protect\citeauthoryear{Zhao, Feng, Shi, and Kakadiaris}{Zhao
  et~al\mbox{.}}{2014}]%
        {statistical-images}
\bibfield{author}{\bibinfo{person}{Xi Zhao}, \bibinfo{person}{Tao Feng},
  \bibinfo{person}{Weidong Shi}, {and} \bibinfo{person}{Ioannis~A.
  Kakadiaris}.} \bibinfo{year}{2014}\natexlab{}.
\newblock \showarticletitle{Mobile User Authentication Using Statistical Touch
  Dynamics Images}.
\newblock \bibinfo{journal}{\emph{IEEE Transactions on Information Forensics
  and Security}} \bibinfo{volume}{9}, \bibinfo{number}{11}
  (\bibinfo{year}{2014}), \bibinfo{pages}{1780--1789}.
\newblock
\urldef\tempurl%
\url{https://doi.org/10.1109/TIFS.2014.2350916}
\showDOI{\tempurl}


\bibitem[\protect\citeauthoryear{Zheng, Bai, Huang, and Wang}{Zheng
  et~al\mbox{.}}{2014}]%
        {tapping}
\bibfield{author}{\bibinfo{person}{Nan Zheng}, \bibinfo{person}{Kun Bai},
  \bibinfo{person}{Hai Huang}, {and} \bibinfo{person}{Haining Wang}.}
  \bibinfo{year}{2014}\natexlab{}.
\newblock \showarticletitle{You Are How You Touch: User Verification on
  Smartphones via Tapping Behaviors}. In \bibinfo{booktitle}{\emph{2014 IEEE
  22nd International Conference on Network Protocols}}.
  \bibinfo{pages}{221--232}.
\newblock
\urldef\tempurl%
\url{https://doi.org/10.1109/ICNP.2014.43}
\showDOI{\tempurl}


\end{thebibliography}
